%% file: finalmain.tex
\documentclass[sigconf]{acmart}

\usepackage{booktabs} 
\usepackage{bm}
\usepackage{amsmath}
\usepackage{verbatim}
\usepackage{multirow}
\usepackage[caption = false]{subfig}

\renewcommand\footnotetextcopyrightpermission[1]{}
\begin{document}
\title{Domain-to-Domain Translation Model for Recommender System}

\author{Linh Nguyen}
\affiliation{%
  \institution{Tohoku University}
}
\email{linh.nguyen.1992@gmail.com}

\author{Tsukasa Ishigaki}
\affiliation{%
  \institution{Tohoku University}
}
\email{isgk@tohoku.ac.jp}
\renewcommand{\shortauthors}{L. Nguyen and T. Ishigaki}

\begin{abstract}
Recently multi-domain recommender systems have received much attention from researchers because they can solve cold-start problem as well as support for cross-selling. However, when applying into multi-domain items, although algorithms specifically addressing a single domain have many difficulties in capturing the specific characteristics of each domain, multi-domain algorithms have less opportunity to obtain similar features among domains. Because both similarities and differences exist among domains, multi-domain models must capture both to achieve good performance. Other studies of multi-domain systems merely transfer knowledge from the source domain to the target domain, so the source domain usually comes from external factors such as the search query or social network, which is sometimes impossible to obtain. To handle the two problems, we propose a model that can extract both homogeneous and divergent features among domains and extract data in a domain can support for other domain equally: a so-called Domain-to-Domain Translation Model (D2D-TM). It is based on generative adversarial networks (GANs), Variational Autoencoders (VAEs), and Cycle-Consistency (CC) for weight-sharing. We use the user interaction history of each domain as input and extract latent features through a VAE-GAN-CC network. Experiments underscore the effectiveness of the proposed system over state-of-the-art methods by a large margin.

\end{abstract}

%
%



\maketitle

\input{finalbody}

\bibliographystyle{ACM-Reference-Format}
\bibliography{references}
\end{document}

%% file: finalbody.tex
\section{Introduction}
\begin{figure*}
\includegraphics[height=2.3in, width=0.8\textwidth]{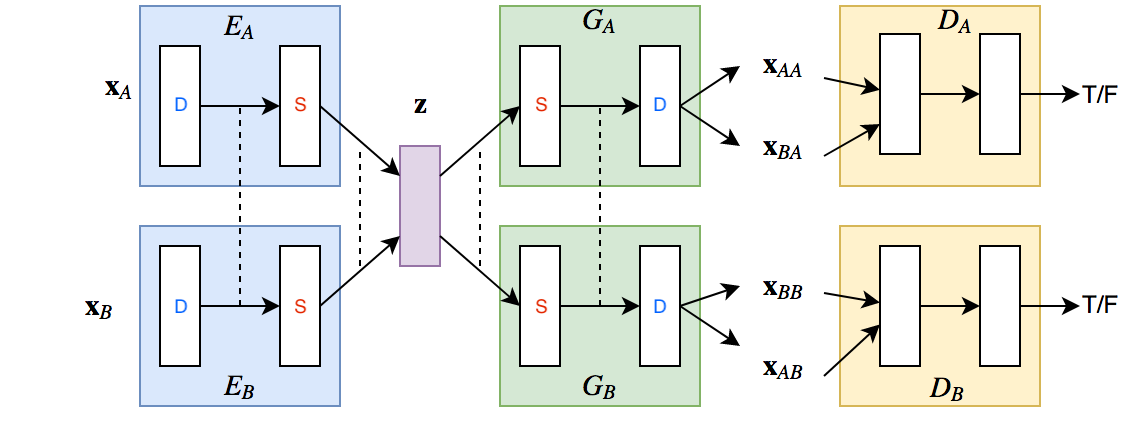}
\caption{We assume that $\mathbf{x}_A$, $\mathbf{x}_B$ are click vectors in two domains A and B from user $u$. Also, $\mathbf{x}_A$ and $\mathbf{x}_B$ can be mapped to the same latent code $\mathbf{z}$ in a shared-latent space $\mathcal{Z}$. $E_A$ and $E_B$ are two encoding functions mapping click vectors to latent codes. $G_A$ and $G_B$ are two generating functions, mapping latent codes to click vectors. We represent $E_A$, $E_B$, $G_A$, and $G_B$ using fully connected layers and implement the shared latent space assumption using a weight-sharing constraint where the connection weights of the last few layers (high-level layers) in $E_A$ and $E_B$ are tied (shown as dashed lines), and where the connection weights of first few layers (high-level layers) in $G_A$, $G_B$ are tied. Therefore, we can learn and generate different features between two domains from the first encoder layers and last generator layers. We can also learn and generate similar features from the last encoder layers and first generator layers. Here, $\mathbf{x}_{AA}$ and $\mathbf{x}_{BB}$ are self-reconstruction vectors, and $\mathbf{x}_{AB}$ and $\mathbf{x}_{BA}$ are domain-translation vectors. $D_A$, $D_B$ are adversarial discriminators for the respected domains in charge of evaluating whether the translated vectors are realistic.}
\label{d2d}
\end{figure*}

During the era of the internet explosion, Recommender Systems (RSs) assumed an important role of supporting users as they find items they need or items reaching the right users. Items can be various products, information, or people, so the RS tends to divide them into small domains in which items have similar attributes \cite{fernandez2012}. Each domain will have specific characteristics. Therefore, to obtain these characteristics, it need to be considered separately. For that reason, many studies have specifically examined a single domain \cite{Gong2016, Chu2017, chen2017attentive}. However, single domains still present numerous difficulties \cite{hu2013personalized}. For example, it can not work well when a user has no interaction in the considered domain or when companies want to cross-sell their products. That these problems are solvable using items from multi-domains \cite{cantador2015cross} leads us to be interested in proposing multi-domain RS.

Algorithms that specifically deal with a single domain can process items from multiple domains easily by aggregating all items into a single domain. However, because all items are learned by a sole network or function, it has difficulties in capturing the specific characteristics of respective domains. On the other hand, some algorithms specifically addressing multiple domains extract latent features of the respective domains by a separated network \cite{Lian2017, min2015cross, }. Although they can highlight the particular features of each domain, they have less opportunity to obtain the similar features among domains. Nevertheless, similarities and differences exist among domains. For that reason, multi-domain systems must capture both to achieve good performance.


In addition, some other multi-domain studies specifically examine the transfer of knowledge from a source domain that is much denser to a target domain, or from specific sources such as user search query or social network information \cite{shapira2013facebook, pan2016mixed, Elkahky2015}. With the first, it is one-way direction. Knowledge of the target domain seems not to be helpful with the source domain. Moreover, many companies are unable to implement the second one because it is sometimes impossible to get these external data.

To address these problems, we propose a multi-domain network structure that can capture both similar and different features among domains and treat every domain equally by taking only implicit feedback inside the system as input. Our model is extended from Unsupervised Image-to-Image Translation Networks (UNIT) \cite{Ming-Yu2017} for the Recommender System, called a Domain-to-Domain Translation Model (D2D-TM). It is based on generative adversarial networks (GANs), Variational Autoencoders (VAEs), and Cycle-Consistency (CC) for weight-sharing. We use the user interaction history of each domain as input, and extract its features through a VAE-GAN-CC network. 

In summary, the main contributions of this paper are the following.
\begin{itemize}
\item Propose a multi-domain recommender system that can extract both homogeneous and divergent features among domains.
\item Translate one domain to another and vice versa simultaneously.
\item Propose an end-to-end deep learning approach for collaborative filtering recommender system which only uses the user interaction history as input
\item Conduct rigorous experiments using two real-world datasets with four couple domains and underscore the effectiveness of proposed system over state-of-the-art methods by a large margin.
\end{itemize}

The remainder of this paper is organized as the following. First, in Section 2, we review related approaches and techniques for recommender systems, which included VAEs, GANs, and a cross-domain recommender system. Section 3 presents an explanation of details of our methods, followed by experiments described in Section 4. We also present conclusions in Section 5.

\section{Related Work}
Extensive studies have been conducted of recommender systems, with reports presented in a myriad of publications. In this section, we aim at reviewing a representative set of approaches that are closely related to our research.
\subsection{Autoencoders }
Autoencoders (AEs) use unsupervised learning which has been shown to be effective for learning latent variables in many deep-learning problems. Collaborative Deep Learning (CDL) \cite{Wang2015} and Collaborative Variational Autoencoder (CVAE) \cite{Li2017} are two well known papers that respectively described application of Denoising Autoencoder and Variational Autoencoder in hybrid methods. Two studies have used Autoencoders to extract latent features from item description text, with related reports proposing joint learning between these latent features and collaborative filtering. Otherwise, the recent method, Multi-VAE \cite{Liang2018} uses VAE to reconstruct user-item matrix to achieve a good result although only using rating information. 

\subsection{Generative Adversarial Networks (GANs)}
As new unsupervised learning networks, GANs can achieve promising results, especially in the realm of computer vision. Nevertheless, few GAN applications have been reported in recommender systems. Actually, IRGAN \cite{Wang2017} is the first model to apply a GAN not only to an information retrieval area but also to a recommender system. IRGAN extends discriminator and generator process in traditional GANs to discriminative retrieval and generative retrieval. Whereas discriminative retrieval learns to predict relevant score $r$ given labeled relevant query-document fairs, generative retrieval tries to generate a fake document to deceive discriminative retrieval. 

Recently, Adversarial Personal Ranking (APR) \cite{He2018}, which enhances the Bayesian personal ranking with adversarial network, and GAN-HBNR \cite{Cai2018}, which proposed a GAN based representation learning approach for heterogeneous bibliographic network, have arisen as new approaches of GAN to recommender systems. 

\begin{figure}
\includegraphics[scale=.2]{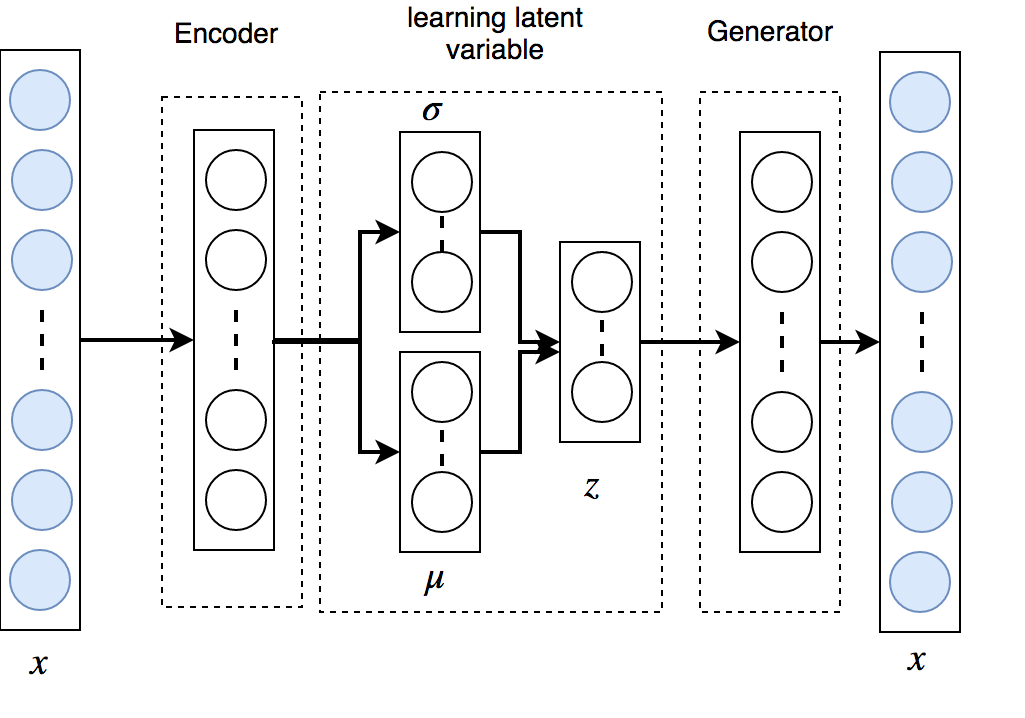}
\caption{General Deep Learning Structure of VAE.}
\label{VAE}
\end{figure}

\subsection{Cross-domain Recommender System}
Today, companies are striving to provide a diversity of products or services to users. For example, Amazon is not only e-commerce platform; it is also online movie and music platform. Therefore, cross-domain recommender systems are necessary for them. Moreover, cross-domain RSs can solve data sparsity and the cold start problem, which are important issues related to single domain RSs. Several works exploring cross-domain RSs have included Multiview Deep Neural Network (MV-DNN) \cite{Elkahky2015}, Neural Social Collaborative Ranking (NSCR) \cite{Wang2017CD}, and Cross-domain Content-boosted Collaborative Filtering neural NETwork (CCCFNET) \cite{Lian2017}. Actually, MV-DNN extracts rich features from the user's browsing and search histories to model a user's interests, whereas item features are extracted from three sources including the title, categories, and contents with news or description with Apps. Then it calculates a relevant score using a cosine function. NSCR attempts to learn embedding of bridge users with user--user connections taken from social networks and user--item interaction. CCCFNET aims to learn content-based embedding so that the model can transfer both content-based and collaborative filtering across different domains simultaneously. A similarity among these methods is that they require external information from other sources. For example, MV-DNN requires a user search query, NSCR combines with user social network account, whereas CCCFNET takes content information. Sometimes, it is impossible to get this knowledge. Therefore, we propose a cross-domain model that uses only implicit feedback inside the system.

\section{Method}
We use $u \in \{1, \cdots, U\}$ to index users, $i_A \in \{1, \cdots, I_A\}$ to index items belonging to domain A, and $i_B \in \{1, \cdots, I_B\}$ to index items belonging to domain B. In this work, we consider learning implicit feedback. The user-by-item interaction matrix is the click \footnote{We use the verb "click" for concreteness. In fact, this can be any type of interaction such as "watch", "view" and "rating"} matrix $\mathbf{X} \in \mathbb{N}^{U \times I}$. The lower case $\mathbf{x}_u = [x_{u1}, x_{u2}, \cdots, x_{uI}]^T \in \mathbb{N}^I$ is a bag-of-words vector, which is the number of clicks for each item from user $u$. With two domains, we have matrix  $\mathbf{X}_A \in \mathbb{N}^{U \times I_A}$ with $\mathbf{x}_{A} = [x_{1A}, x_{2A}, \cdots, x_{IA}]^T \in \mathbb{N}^{I_A}$ for domain A, and $\mathbf{X}_B \in \mathbb{N}^{U \times I_B}$ with $\mathbf{x}_{B} = [x_{1B}, x_{2B}, \cdots, x_{IB}]^T \in \mathbb{N}^{I_B}$ for domain B. For simplicity, we binarize the click matrix. It is straightforward to extend it to general count data.

\subsection{Framework}
Our framework, as presented in Figure \ref{d2d}, is based on variational autoencoder (VAE) and generative adversarial network (GAN). It comprises six subnetworks including two domain click vector encoders $E_A$ and $E_B$, two domain click vector generators $G_A$ and $G_B$, and two domain adversarial discriminators $D_A$ and $D_B$. We maintain the framework structure as \cite{Ming-Yu2017}. We share weights of the last few layers in $E_A$ and $E_B$, so that our model not only extracts different characteristics of two model in the first layers, but also learns their similarity. In parallel, we also share weights of the few first layers in $G_A$ and $G_B$ to make our model able to generate both similar and divergent features. In Figure \ref{d2d}, share layers are denoted as {\color{red}S}, whereas distinct layers are denoted as {\color{blue}D}. Moreover, the GAN model constraint generated vectors of two domains distinctly. Otherwise, our framework learns translation in both directions in one shot.
\subsection{VAE}
Figure \ref{VAE} portrays the general structure of VAE. In our model, the encoder--generator pair $\{E_A, G_A\}$ constitutes a VAE for domain A, term $\text{VAE}_A$. For an input click vector $\mathbf{x}_A \in A$, the $\text{VAE}_A$ first maps $\mathbf{x}_A$ to a code in a latent space $\mathcal{Z}$ via encoder $E_A$. It then decodes a randomly perturbed version of the code to reconstruct the input click vector via the generator $G_A$. We assume that the components in the latent space $\mathcal{Z}$ are conditionally independent and Gaussian with unit variance. In our formulation, the encoder outputs a mean vector $E_{\mu,A}(\mathbf{x})_A$. The distribution of the latent code $\mathbf{z}_A$ is given as $q_A(\mathbf{z}_A|\mathbf{x}_A \equiv \mathcal{N}(\mathbf{z}_A|E_{\mu, A}(\mathbf{x}_A, \mathbf{I}))$, where $\mathbf{I}$ is an identity matrix. The reconstructed click vector is $\mathbf{x}_{AA} = G_A(\mathbf{z}_A \sim q_A(\mathbf{z}_A|\mathbf{x}_A))$.  

Similarly, $\{E_{B}, G_{B}\}$ constitutes a VAE for domain B: $\mathbf{\text{VAE}_B}$ where the encoder $E_B$ outputs a mean vector $E_{\mu, B}(\mathbf{x}_B)$ and the distribution of latent code $\mathbf{z}_B$ is given as $q_B(\mathbf{z}_B|\mathbf{x}_B) \equiv \mathcal{N}(\mathbf{z}_B|E_{\mu, B}(\mathbf{x}_B, \mathbf{I}))$. The reconstructed click vector is $\mathbf{x}_{BB} = G_B(\mathbf{z}_B \sim q_B(\mathbf{z}_B|\mathbf{x}_B))$.

\subsection{Weight-sharing and Cycle-consistency (CC)}
We enforce a weight-sharing constraint relating two VAEs. Specifically, we share the weights of the last few layers of $E_A$ and $E_B$ that are responsible for extracting high-level representations of the input click vectors in the two domains. Similarly, we share the weights of the first few layers of $G_A$ and $G_B$ responsible for decoding high-level representations for reconstructing the input click vector.

The shared latent space assumption enables us to use domain-to-domain translation. We can translate a click vector $\mathbf{x}_A$ in domain A to a click vector in domain B through applying $G_B(\mathbf{z}_A\sim q_A(\mathbf{z}_A | \mathbf{x}_A))$. Similarly, click vector from domain B to domain A is generated as $G_A(\mathbf{z}_B \sim q_B(\mathbf{z}_B | \mathbf{x}_B))$.

We also use cycle-consistency as a weight-sharing constraint.

\subsection{Generative Adversarial Network}
\begin{figure}
\includegraphics[scale=.2]{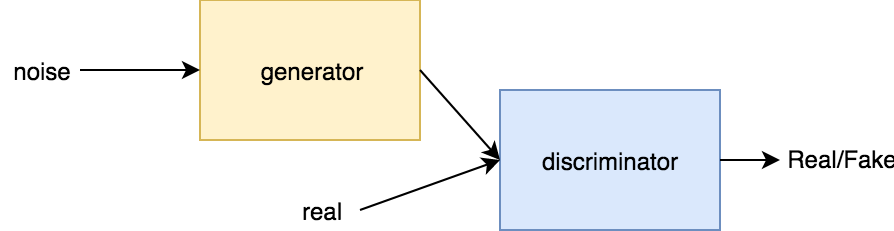}
\caption{General Structure of GAN.}
\label{GAN}
\end{figure}
Figure \ref{GAN} shows a general structure of GAN. Our framework has two generative adversarial networks: $\text{GAN}_A = \{ D_A, G_A \}$ and $\text{GAN}_B = \{ D_B, G_B \}$. In $\text{GAN}_A$, for real click vectors sampled from the first domain, $D_A$ should output true, although it should output false for a click vector generated by $G_A$. 

Click vectors of two types can be generated from $G_A$.

$\mathbf{x}_{AA} = G_A(\mathbf{z}_A \sim q_A(\mathbf{z}_A|\mathbf{x}_A)), \text{and,}$

$\mathbf{x}_{BA} = G_A(\mathbf{z_B} \sim q_B(\mathbf{z}_B |\mathbf{x}_B))$

Because the reconstruction stream can be trained supervisedly, it is sufficient that we merely apply adversarial training to click vectors from the translation stream $\mathbf{x}_{BA}$.

We apply a similar processing to $\text{GAN}_B$, where $D_B$ is trained to output true for real click vectors sampled from the second domain dataset and to output false for click vectors generated from $G_B$

\subsection{Learning}
We solve the learning problems of $\text{VAE}_A$, $\text{VAE}_B$, $\text{GAN}_A$, and $\text{GAN}_B$ jointly.
\begin{align}
\displaystyle \min_{E_A, E_B, G_A, G_B} \max_{D_A, D_B} &\mathcal{L}_{\text{VAE}_A} (E_A, G_A) + \mathcal{L}_{\text{GAN}_A}(E_B, G_A, D_A) \nonumber \\
& + \mathcal{L}_{\text{CC}_A}(E_A, G_A, E_B, G_B) \nonumber\\
& \mathcal{L}_{\text{VAE}_B} (E_B, G_B) + \mathcal{L}_{\text{GAN}_B}(E_A, G_B, D_B) \nonumber\\
& + \mathcal{L}_{\text{CC}_B}(E_B, G_B, E_A, G_A) 
\label{learning}
\end{align}

\subsubsection{VAE}: VAE training aims for minimizing a variational upper bound. In (\ref{learning}), the VAE objects are the following.

\begin{align}
\mathcal{L}_{\text{VAE}_A} & = \lambda_1 \textbf{KL}(q_A(\mathbf{z}_A|\mathbf{x}_A)\|p_\eta (\mathbf{z})) - \lambda_2\mathbb{E}_{\mathbf{z}_A \sim q_A(\mathbf{z}_A|\mathbf{x}_A)}[\log p_{G_A}(\mathbf{x}_A|\mathbf{z}_A)] \\
\mathcal{L}_{\text{VAE}_B} & = \lambda_1 \textbf{KL}(q_B(\mathbf{z}_B|\mathbf{x}_B)\|p_\eta (\mathbf{z})) - \lambda_2\mathbb{E}_{\mathbf{z}_B \sim q_A(\mathbf{z}_B|\mathbf{x}_B)}[\log p_{G_B}(\mathbf{x}_B|\mathbf{z}_B)] 
\end{align}

Therein, the hyperparameters $\lambda_1$ and $\lambda_2$ control the weights of the objective terms and the \textbf{KL} divergence terms penalize deviation of the distribution of the latent code from the prior distribution. The regularization allows an easy means of sampling from the latent space. The prior distribution is a zero mean Gaussian $p_\eta (\mathbf{z}) = \mathcal{N}(\mathbf{z}|0, \mathbb{I})$.

We model $p_{G_A}$ and $p_{G_B}$ as paper \cite{Liang2018}. Therefore, with a user, minimizing log-likelihood term is equivalent to minimizing the multinomial likelihood for click vector as

\begin{align*}
\log p_{G_A}(\mathbf{x}_A|\mathbf{z}_A) &= \sum_i^{I_A} \mathbf{x}_{i,A}\log f  (\mathbf{x}_{i,AA}) \\
\log p_{G_B}(\mathbf{x}_B|\mathbf{z}_B) &= \sum_i^{I_B} \mathbf{x}_{i,B}\log f (\mathbf{x}_{i,BB}),
\end{align*}
where $f(.)$ is a softmax function.

\subsubsection{GAN}: In (\ref{learning}), the GAN objective functions are given as shown below.
\begin{align}
\mathcal{L}_{\text{GAN}_A}(E_B, G_A, D_A) &= \lambda_0\mathbb{E}_{\mathbf{x}_A\sim P_A[\log D_A(\mathbf{x}_A)]} + \nonumber \\
& \lambda_0\mathbb{E}_{\mathbf{z}_B\sim q_B(\mathbf{z}_B|\mathbf{x}_B)}[\log(1-D_A(G_A(\mathbf{z}_B)))] \label{gan1} \\
\mathcal{L}_{\text{GAN}_B}(E_A, G_B, D_B) &= \lambda_0\mathbb{E}_{\mathbf{x}_B\sim P_B[\log D_B(\mathbf{x}_B)]} + \nonumber\\
& \lambda_0\mathbb{E}_{\mathbf{z}_A\sim q_A(\mathbf{z}_A|\mathbf{x}_A)}[\log(1-D_B(G_B(\mathbf{z}_A)))] \label{gan2}
\end{align}

The objective functions in (\ref{gan1}) and (\ref{gan2}) are conditional GAN objective functions. They are used to ensure the translated click vectors resembling click vectors in target domains. Hyperparameter $\lambda_0$ controls the effect of the GAN objective functions.

\subsubsection{CC}: We use a VAE-like objective function to model the cycle-consistency constraint, which is given as presented below.

\begin{align}
\mathbf{L}_{CC_A} (E_A, G_A, E_B, G_B) = &\lambda_3\textbf{KL}(q_A(\mathbf{z}_A|\mathbf{x}_A)\| p_\eta(\mathbf{z})) + \nonumber \\
& \lambda_3\textbf{KL}(q_B(\mathbf{z}_B|\mathbf{x}_{AB})\| p_\eta(\mathbf{z})) - \nonumber\\
& \lambda_4\mathbb{E}_{\mathbf{z}_B \sim q_A(\mathbf{z}_B|\mathbf{x}_{AB})}[\log p_{G_A}(\mathbf{x}_A|\mathbf{z}_B)] \\
\mathbf{L}_{CC_B} (E_B, G_B, E_A, G_A) = &\lambda_3\textbf{KL}(q_B(\mathbf{z}_B|\mathbf{x}_B)\| p_\eta(\mathbf{z})) + \nonumber \\
& \lambda_3\textbf{KL}(q_A(\mathbf{z}_A|\mathbf{x}_{BA})\| p_\eta(\mathbf{z})) - \nonumber\\
& \lambda_4\mathbb{E}_{\mathbf{z}_A \sim q_B(\mathbf{z}_A|\mathbf{x}_{BA})}[\log p_{G_B}(\mathbf{x}_B|\mathbf{z}_A)] 
\end{align} 
Therein, the negative log-likelihood objective term ensures that a twice-translated click vector resembles the input one and that the \textbf{KL} terms penalize the latent codes deviating from the prior distribution in the cycle-reconstruction stream. Hyperparameters $\lambda_3$ and $\lambda_4$ control the weights of the two objective terms.

\subsection{Predict}
\subsubsection{From domain A to domain B}
Assuming a history click vector $\mathbf{x}_A$ of user $u$ in domain A, we predict click vector $\mathbf{x}_{AB}$ of user $u$ in domain B as $\mathbf{x}_{AB} \sim G_B(E_A(\mathbf{x}_A))$.

\subsubsection{From domain B to domain A}
Similarly, with a history click vector $\mathbf{x}_B$ of user $u$ in domain B, we predict his click vector $\mathbf{x}_{BA}$ in domain A as $\mathbf{x}_{BA} \sim G_A(E_B(\mathbf{x}_B))$.

\section{Experiments}
This section presents an evaluation of our proposed method on real-world datasets of Amazon\footnote{http://jmcauley.ucsd.edu/data/amazon/} and Movielens\footnote{https://grouplens.org/datasets/movielens/}. Then we present a comparison with others state-of-the-art methods. The experimentally obtained results constitute evidence of significant improvement over competitive baselines.

\subsection{Dataset Description}
\begin{table}
\caption{Of datasets after preprocessing, \#user, \#item\_A, and \#item\_B respectively represent the number of users, the number of items in domain A, and the number of item in domain B. Dense\_A and dense\_B respectively refer to the density percentages of rating matrixes from domain A and domain B}.

\begin{tabular}{|l|l|l|l|l|l| }
\hline
  Dataset & \#user & \#item\_A & \#item\_B &dense\_A & dense\_B \\ \hline
  Health\_Clothing & 6557 & 16069 &18226 &0.08 & 0.05 \\ \hline
  Video\_TV & 5459 & 10072 & 28578 & 0.14 & 0.1 \\ \hline
  Drama\_Comedy & 6023 & 1490 & 1081 & 3.3 & 3.3 \\ \hline
  Romance\_Thriller & 5891 & 455 & 475 & 5.27 & 6.4\\ \hline
\end{tabular}
\label{datasetsum}

\end{table}

\subsubsection{Amazon} 
We created two datasets from four Amazon review subsets: Health\_Clothes from Health and Personal Care and Clothing, Shoes and Jewelry; Video\_TV from Video Games and Movies and TV. In each dataset, we maintained a list of users who review in both subsets as well as products which the users reviewed. We treat the rating as implicit feedback.
\[ r_{ij} =
  \begin{cases}
    1      & \quad \text{if user } i \text{ rated for item }j\\
    0  & \quad \text{otherwise } 
  \end{cases}
\]

\subsubsection{Movielens} 
>From dataset Movielens 1M, we created two subsets: Drama\_Comedy and Romance\_Thriller. The Drama\_Comedy dataset included users who rated for both Drama and Comedy movies, Drama movies and Comedy, which the users rated and rating scores among them. We similarly prepared Romance\_Thriller and consider rating scores as implicit feedback of the Amazon dataset.

We named datasets following an A\_B structure. For instance, the dataset designated as Health\_Clothing means domain A is Health and Personal Care products; domain B is Clothing, Shoes and Jewelry products. After preprocessing, we have details of four datasets as in Table \ref{datasetsum}. From this, it is apparent that Movielens is much denser than Amazon. Hence, it can be considered as we tested our model in both sparse and dense case.

\subsection{Evaluation Scheme}
We use two ranking based metrics: Recall@K and normalized discounted cumulative gain (NDCG@K) \cite{wang2013theoretical}. With each user, we sort the predicted list and take the K highest score items. Then we compare with ground truth items. Recall@K is defined as the percentage of purchase items which are of the recommended list:  
\begin{align*}
\text{Recall@K} = \frac{\text{Number of items that a user likes in the top K}}{\text{total number of items that a user likes}} 
\end{align*}

However, NDCG@K is defined as the most frequently used list evaluation measure that incorporates consideration of the position of correctly recommend items. First, we consider the discounted cumulative gain (DCG) of a user as
\begin{align*}
\text{DCG@K} = \sum_{i=1}^K \frac{2^{hit_i} - 1}{\log_2(i+1)}
\end{align*}
where \[ hit_i =
  \begin{cases}
    1      & \quad \text{if item } i^{th} \text{ in groud truth list }\\
    0  & \quad \text{otherwise } 
  \end{cases}
\]
Because DCG is unequal among users, we normalize it as
\begin{align*}
\text{NDCG@K} = \frac{DCG@K}{IDCG@K},
\end{align*}
where ideal discounted cumulative gain is represented as IDCG.
\begin{align*}
\text{IDCG@K} = \sum_{i=1}^{|HIT|} \frac{2^{hit_i} - 1}{\log_2(i+1)}
\end{align*}
Therein, $|HIT|$ is a list of ground truth up to position K.

The final result reported the average over all users.
\subsection{Experimental Settings}
We divided all users in each dataset randomly following 70\% for training, 5\% for validation to optimize hyperparameters, and 25\% for testing. We train models using the entire click history of train users. In validation and test processes, we use a click vector of domain A to predict the click vector of domain B and vice versa. We will choose settings which gave the best recall@50 in validation sets. 

The overall structure for the Drama\_Comedy and Romance\_Thriller dataset is [I-200-100-50-100-200-I], whereas the first [100] is the shared layer in the encoder. The second [100] is the shared layer in the generator, [50] represents the latent vector dimension, and $I$ stands for the number of products in domain A or B. 

For the Amazon dataset, because the number of products in each domain is much greater than in the Movielens dataset, the overall structure for Health\_Clothing and Video\_TV dataset is [I-600-200-50-200-600-I], whereas the first [200] is share-layer in encoder, the second [200] is the share-layer in the generator, [50] is latent vector dimension, and $I$ is number of products in domain A or B. We also found that with sparse dataset as Amazon, adding a dropout layer to the input layer will give a better result.

With each hidden layer in encoder and generator, we apply a leaky ReLU activation function with scale of 0.2. With discriminator network, we apply tanh function for each hidden layer, except the last layer.

\subsection{Baselines}
\begin{figure*}
\subfloat[]{\includegraphics[width = 1.7in]{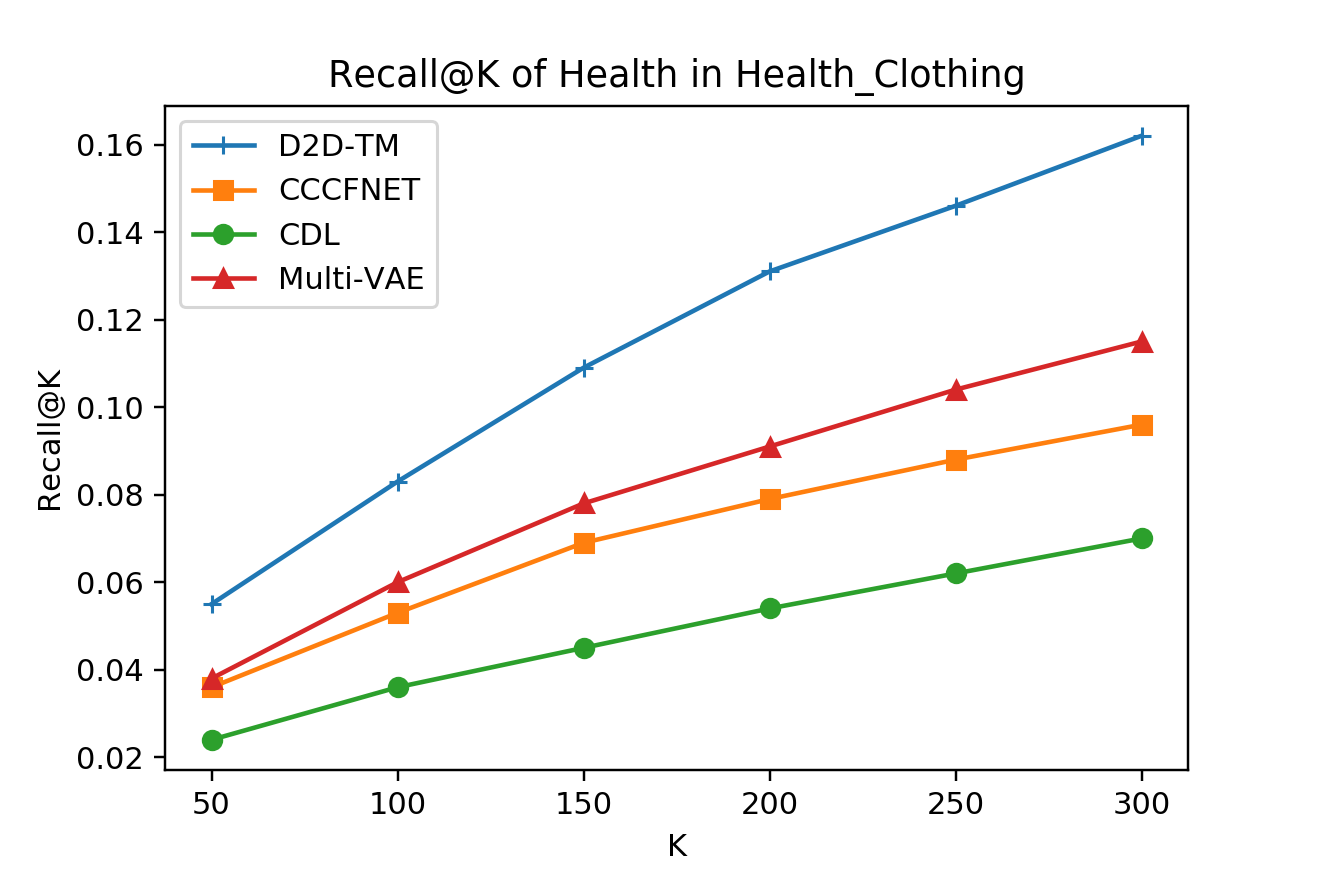}} 
\subfloat[]{\includegraphics[width = 1.7in]{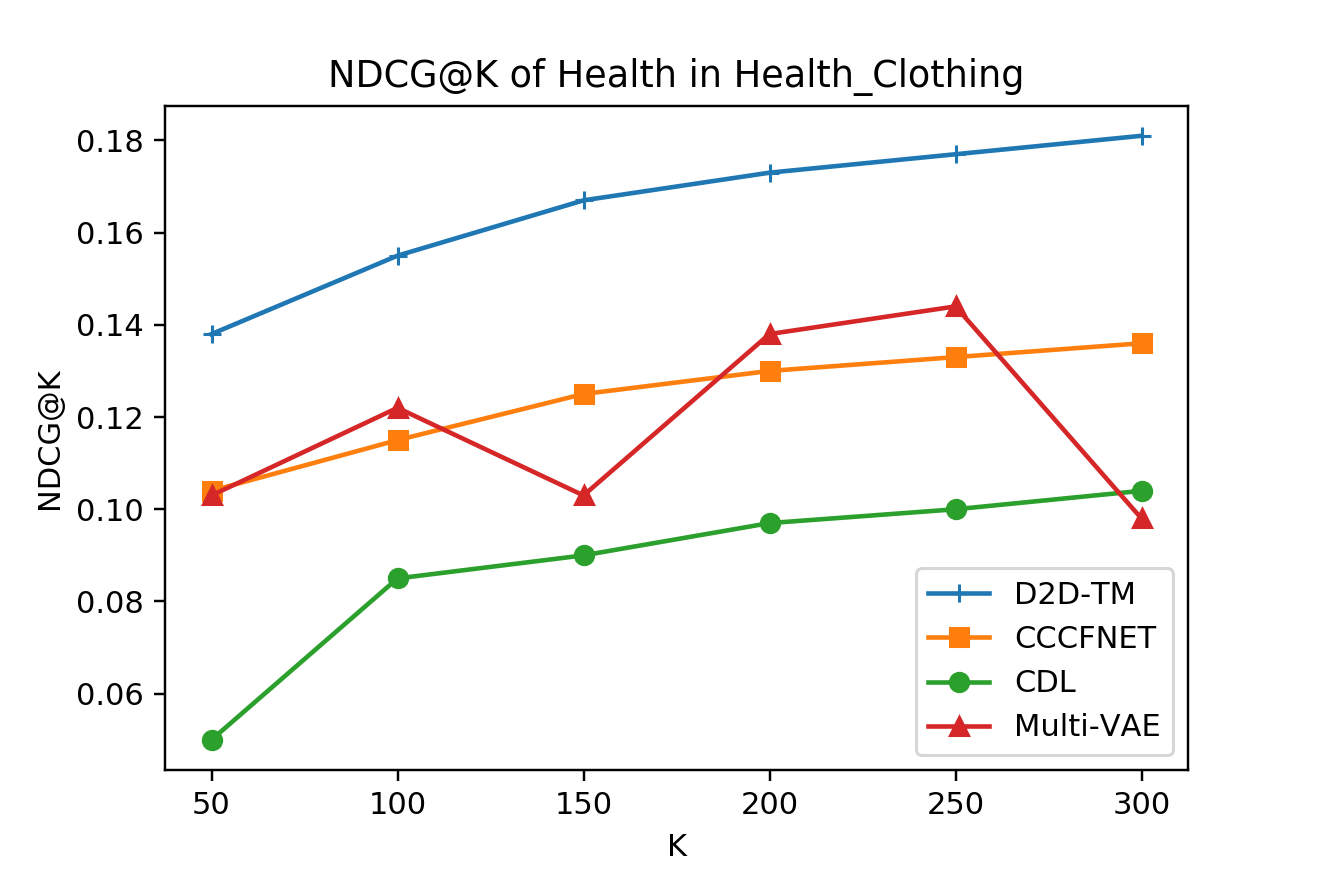}}
\subfloat[]{\includegraphics[width = 1.7in]{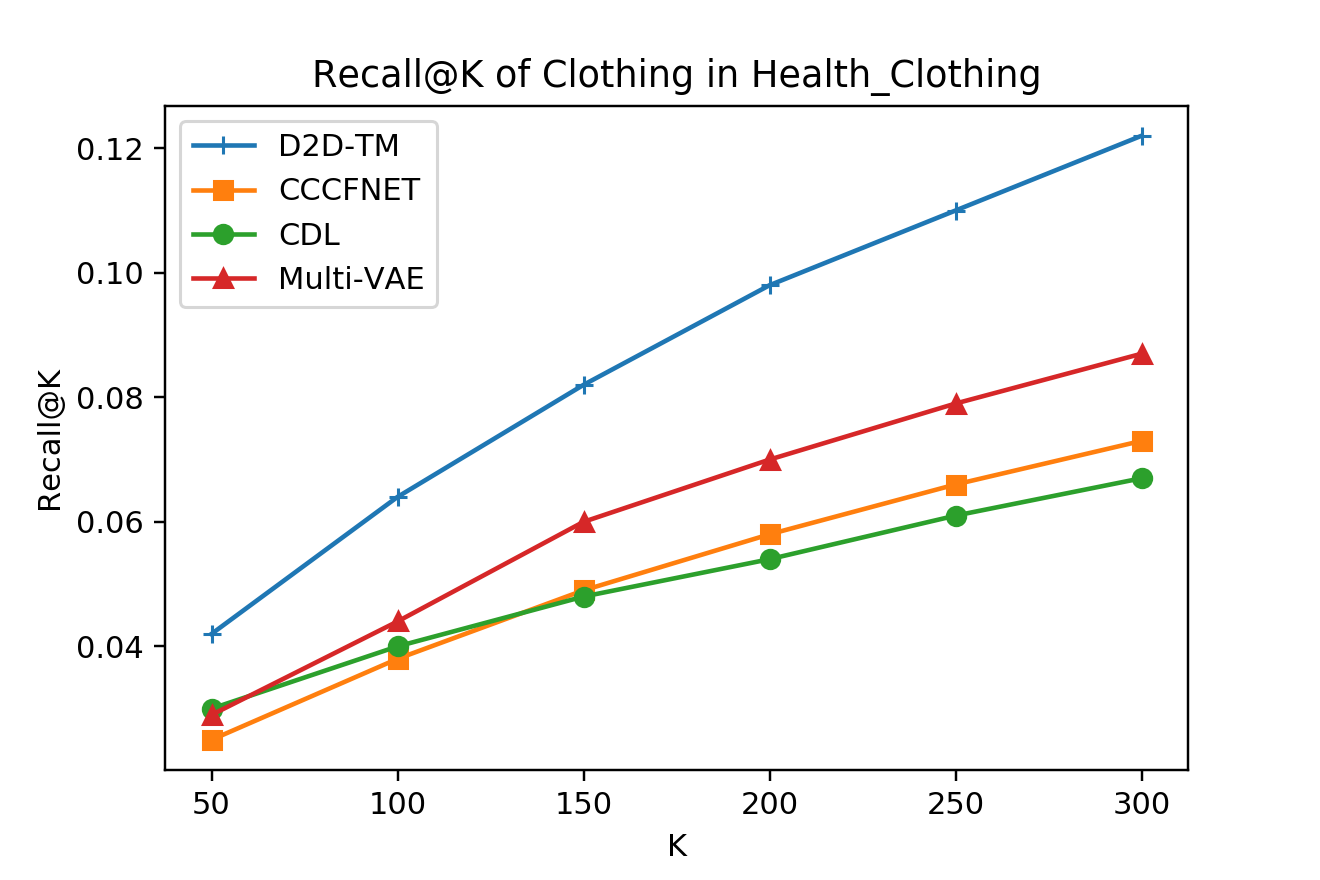}}
\subfloat[]{\includegraphics[width = 1.7in]{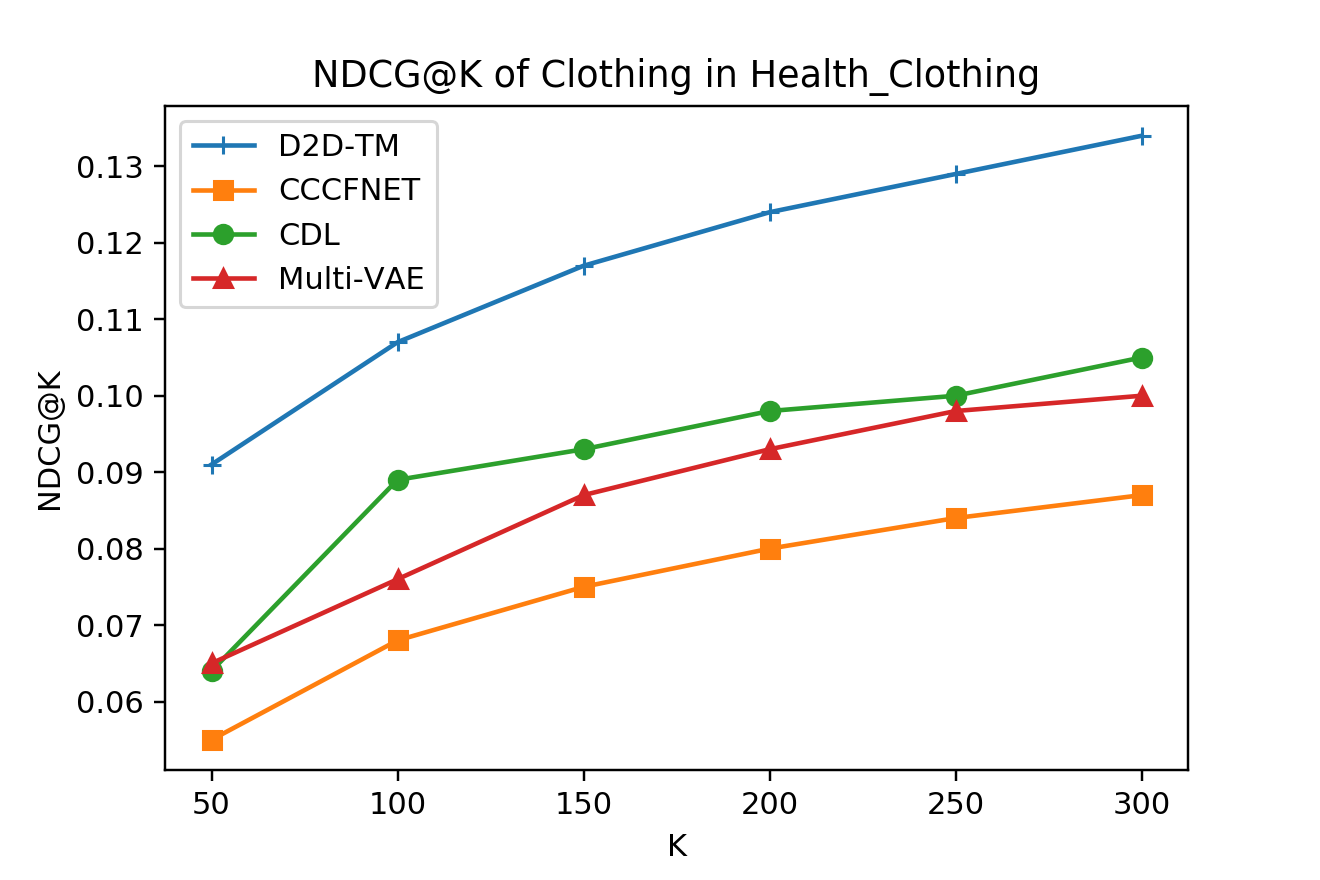}} \\
\subfloat[]{\includegraphics[width = 1.7in]{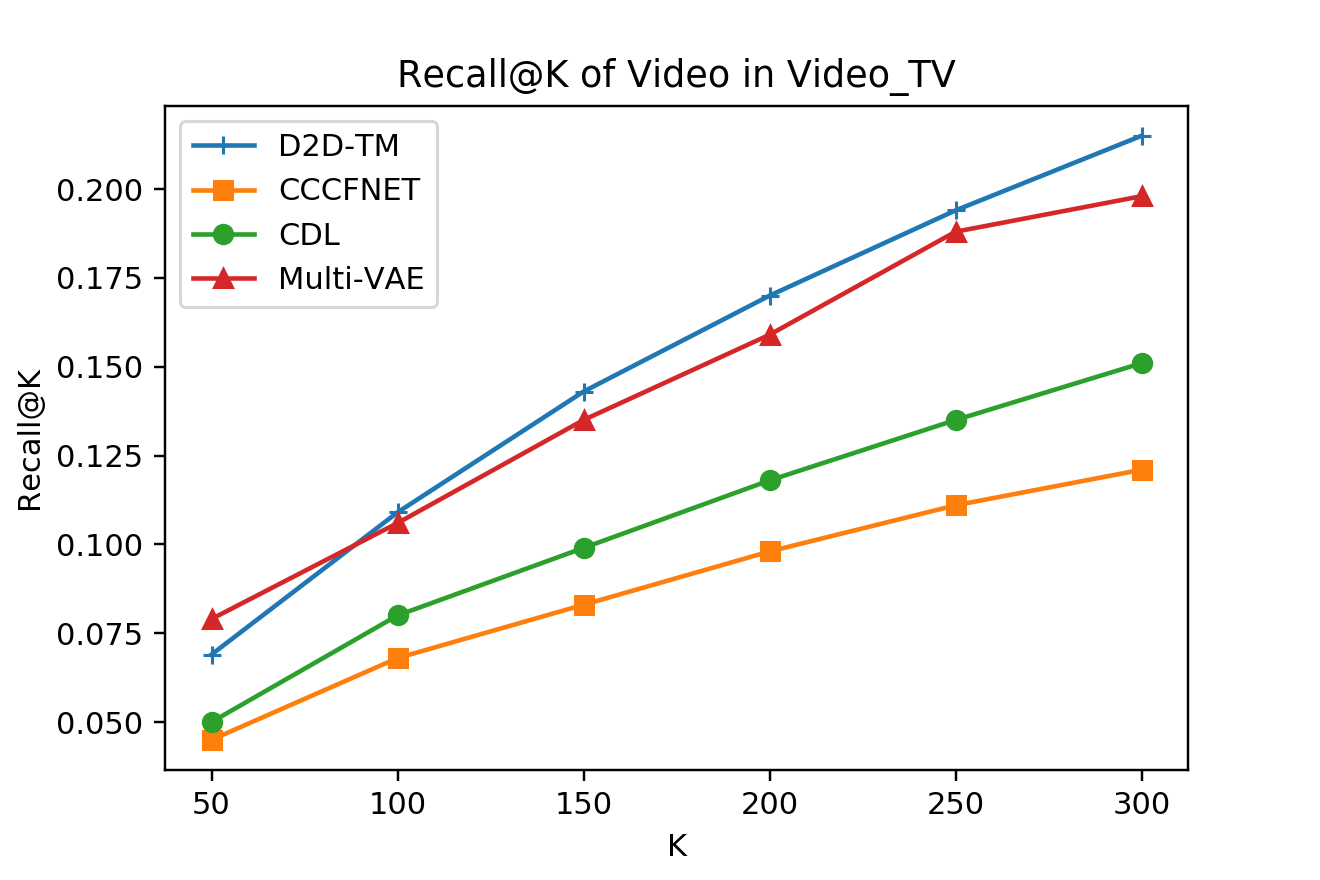}}
\subfloat[]{\includegraphics[width = 1.7in]{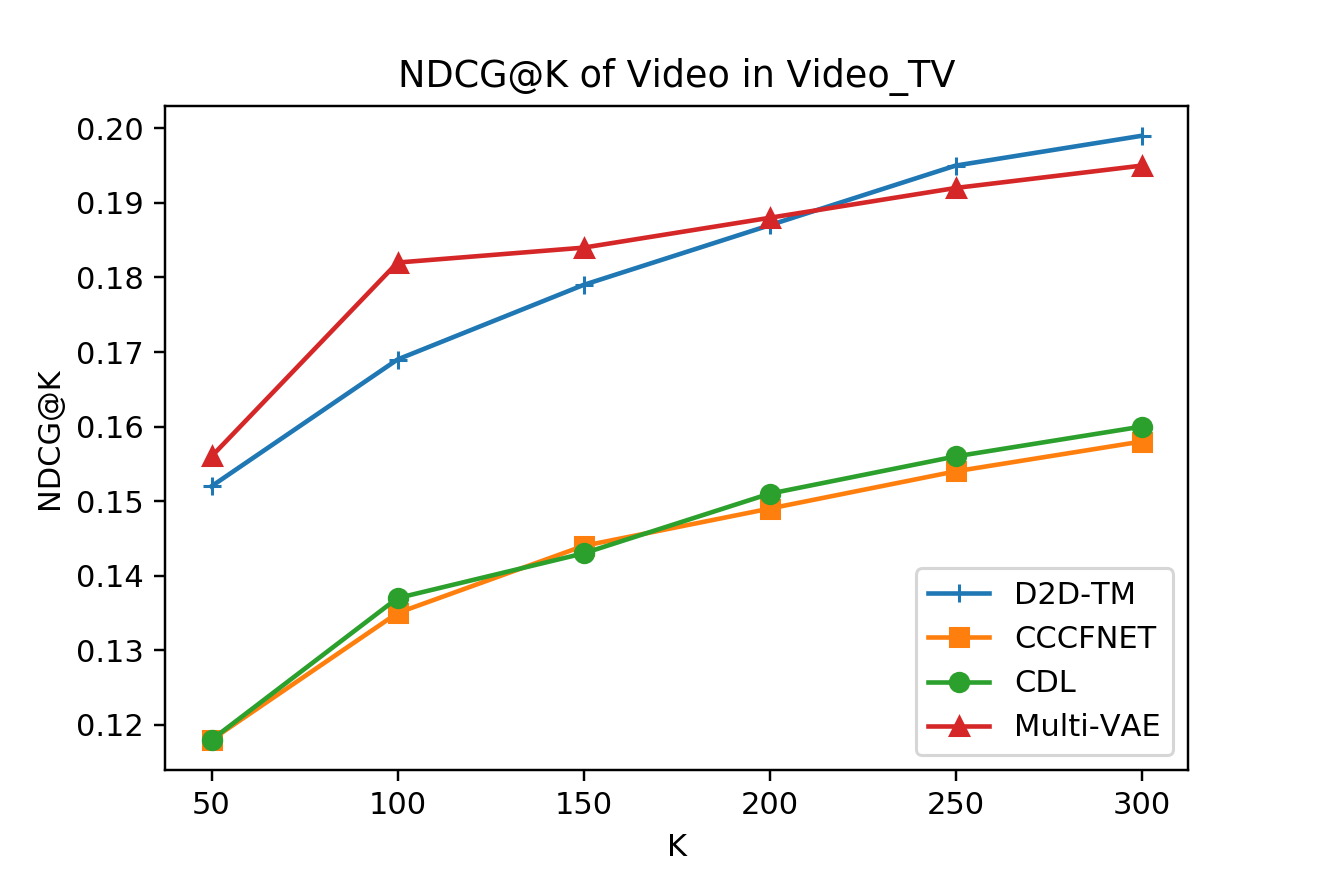}}
\subfloat[]{\includegraphics[width = 1.7in]{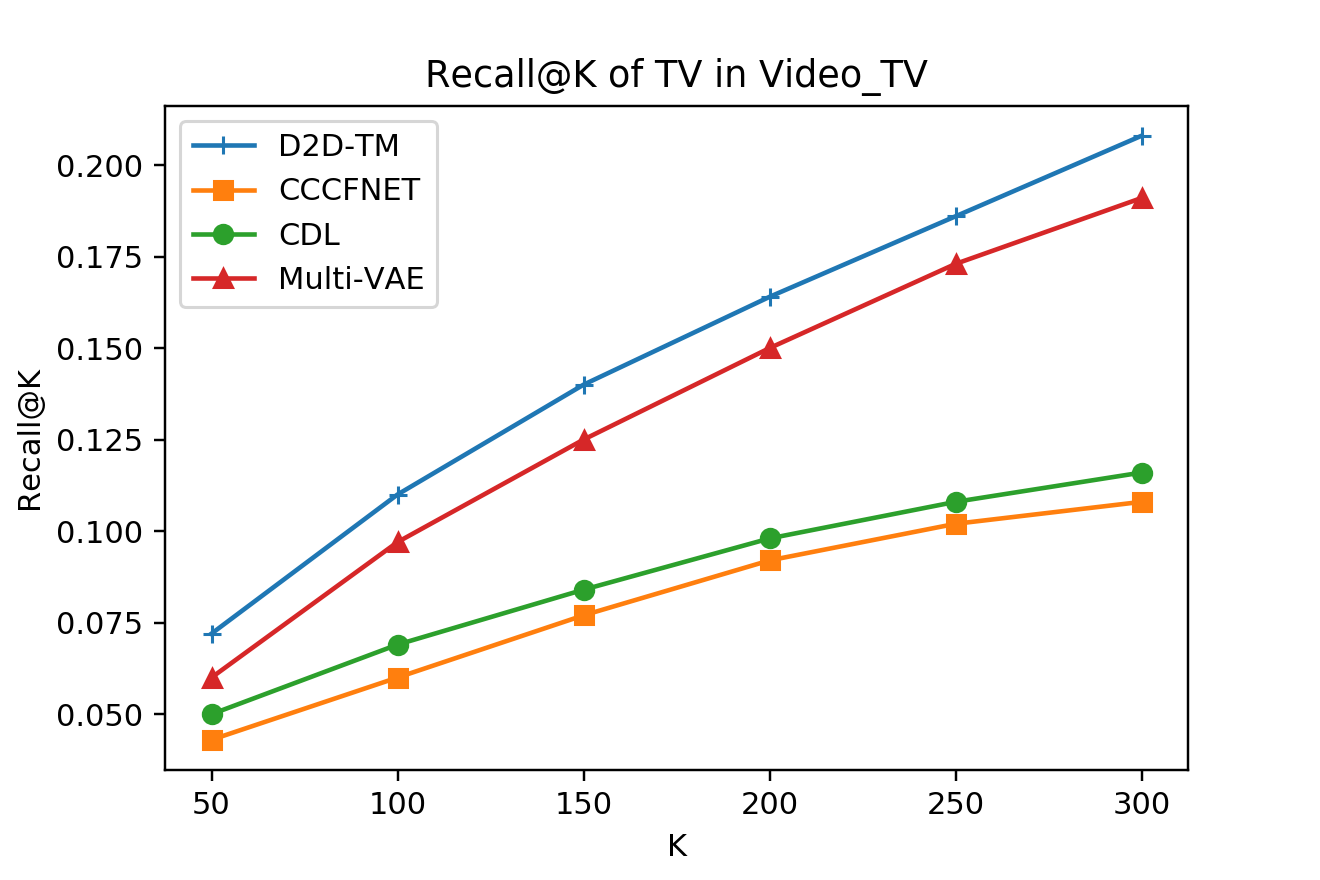}}
\subfloat[]{\includegraphics[width = 1.7in]{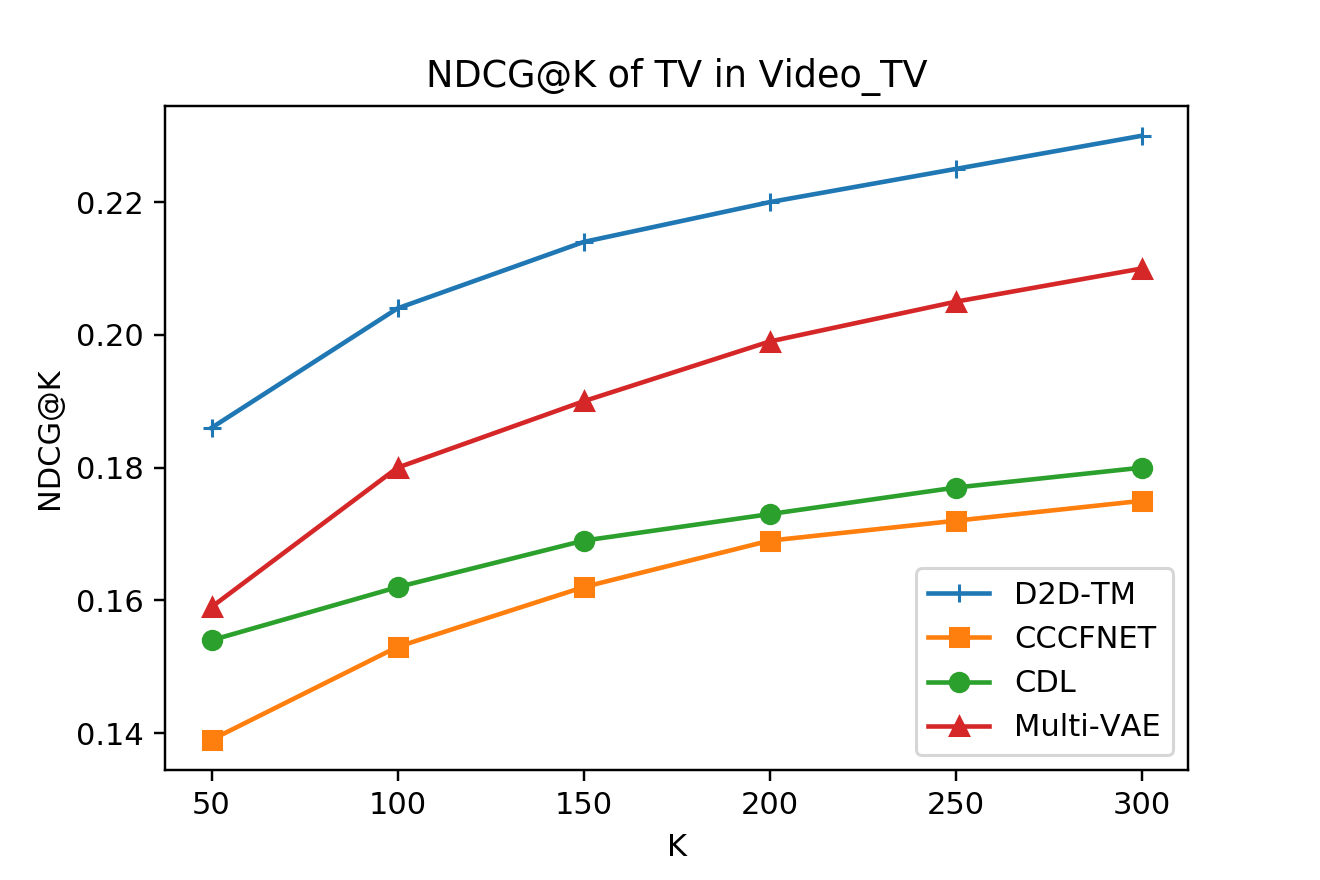}} \\
\subfloat[]{\includegraphics[width = 1.7in]{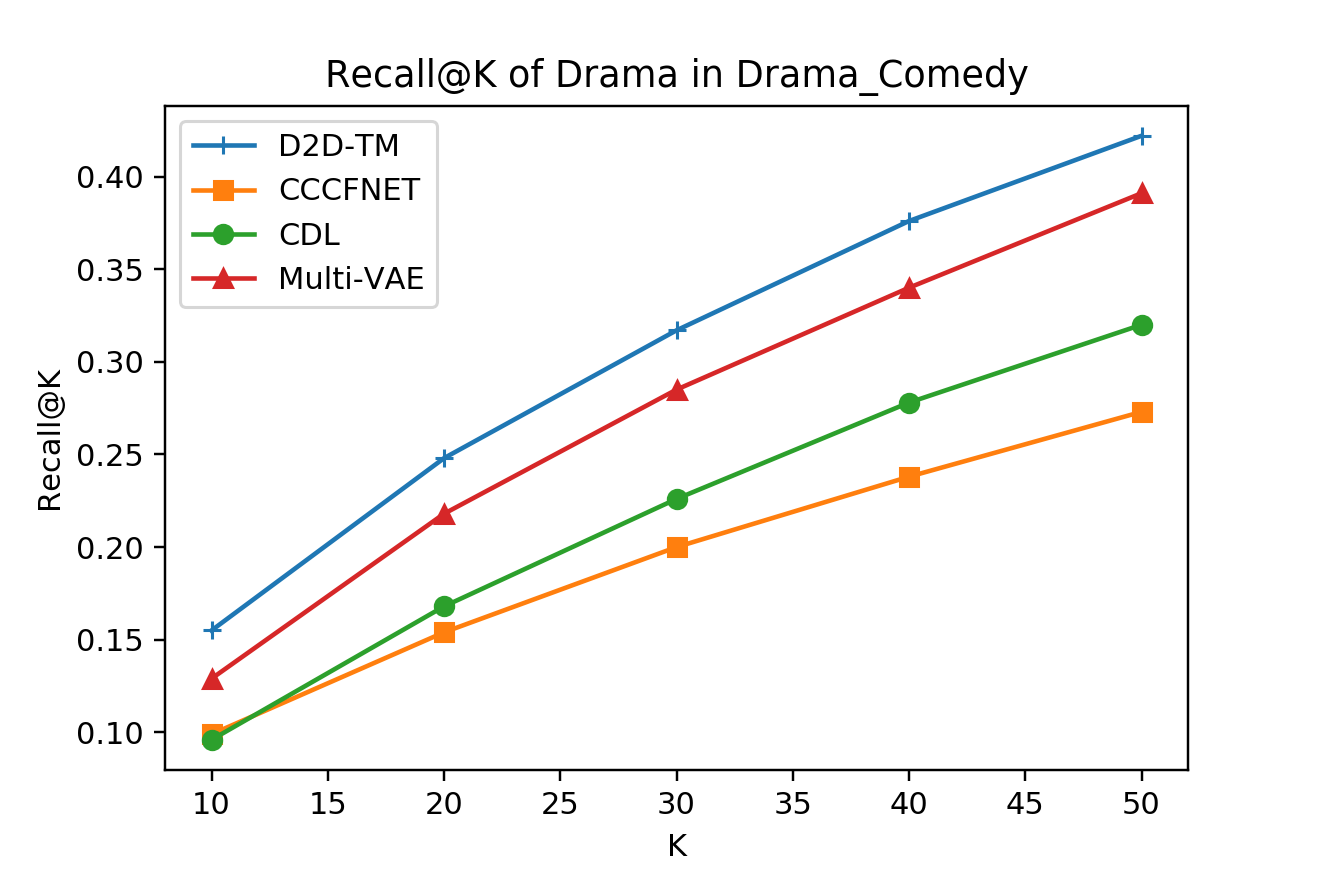}}
\subfloat[]{\includegraphics[width = 1.7in]{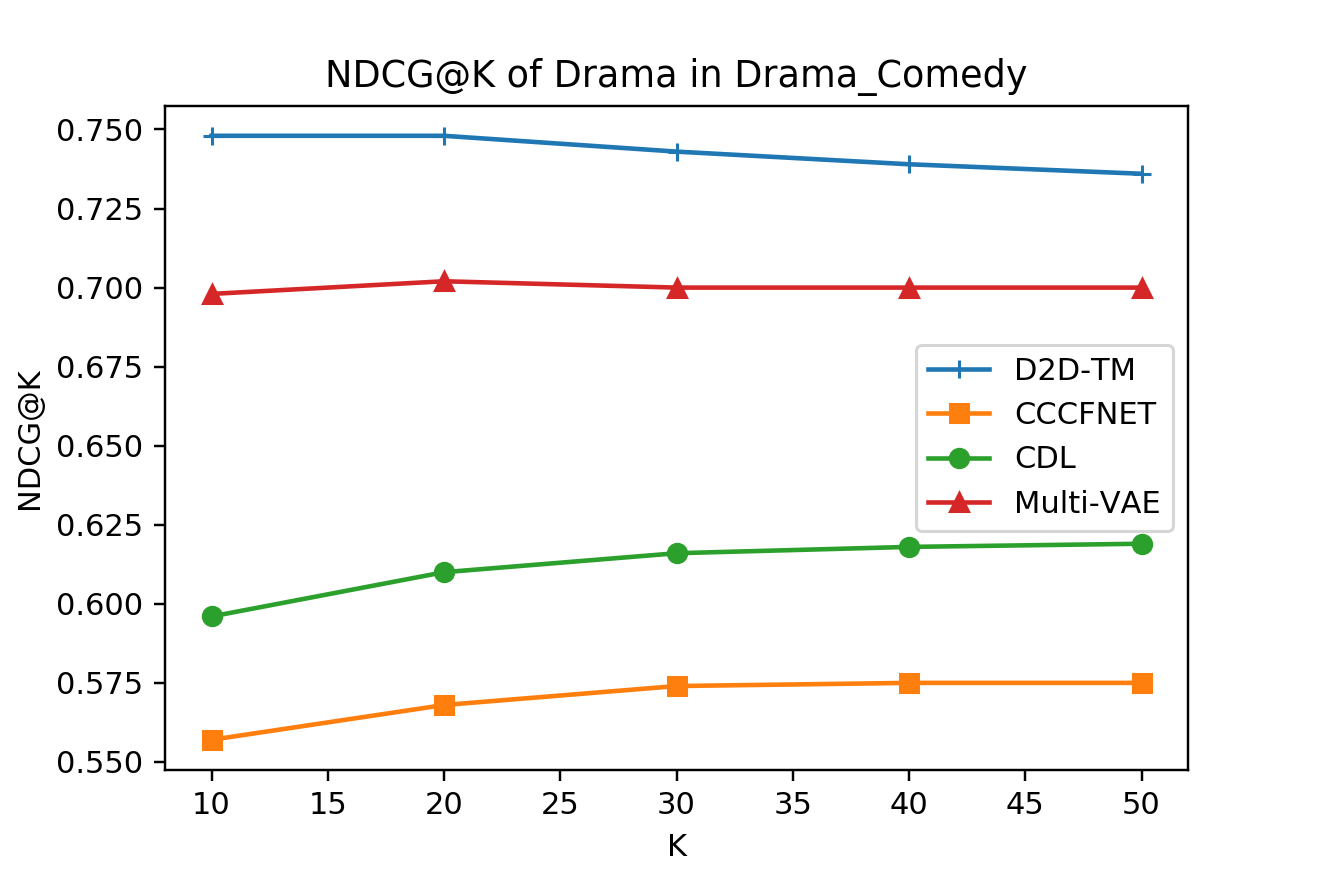}}
\subfloat[]{\includegraphics[width = 1.7in]{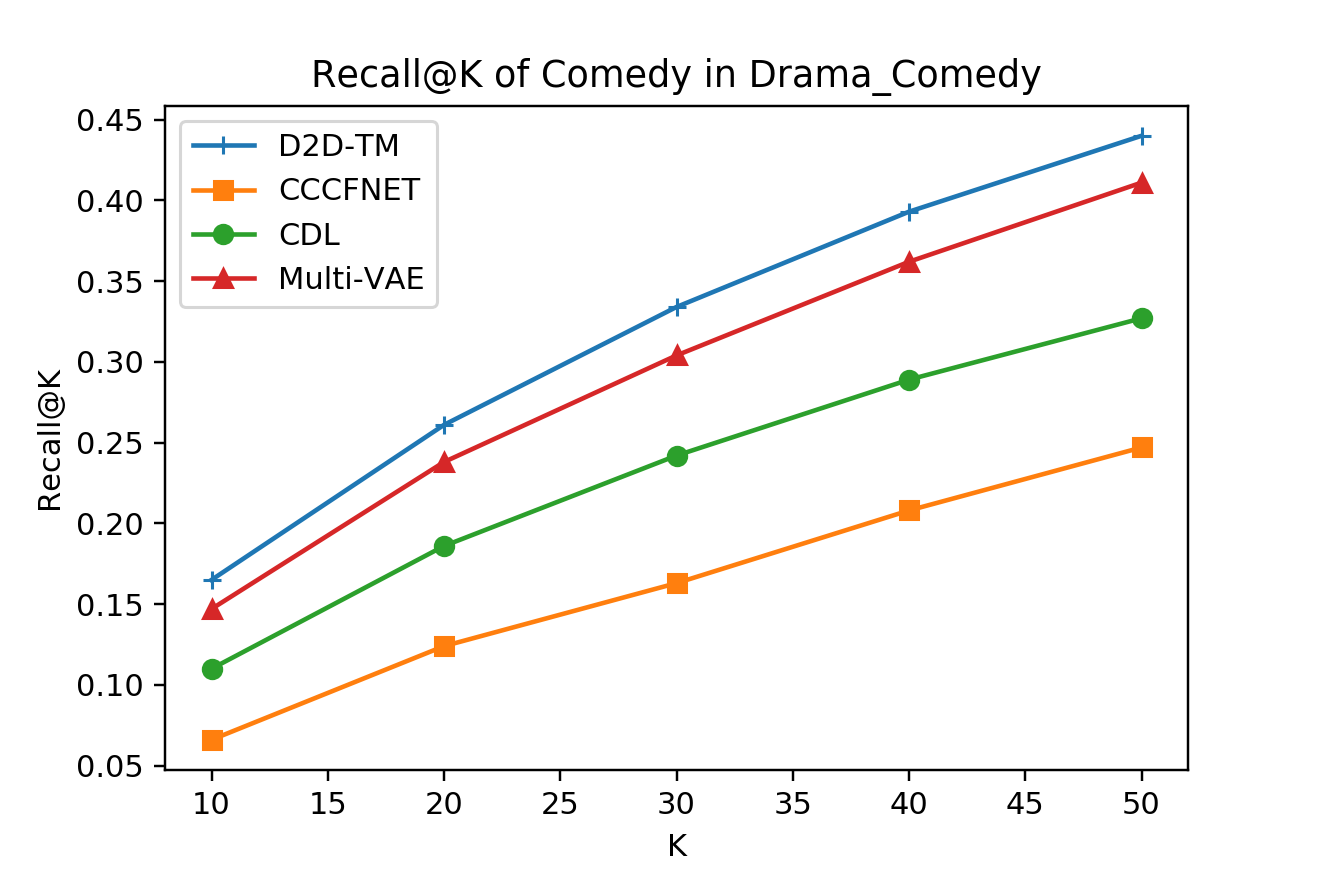}}
\subfloat[]{\includegraphics[width = 1.7in]{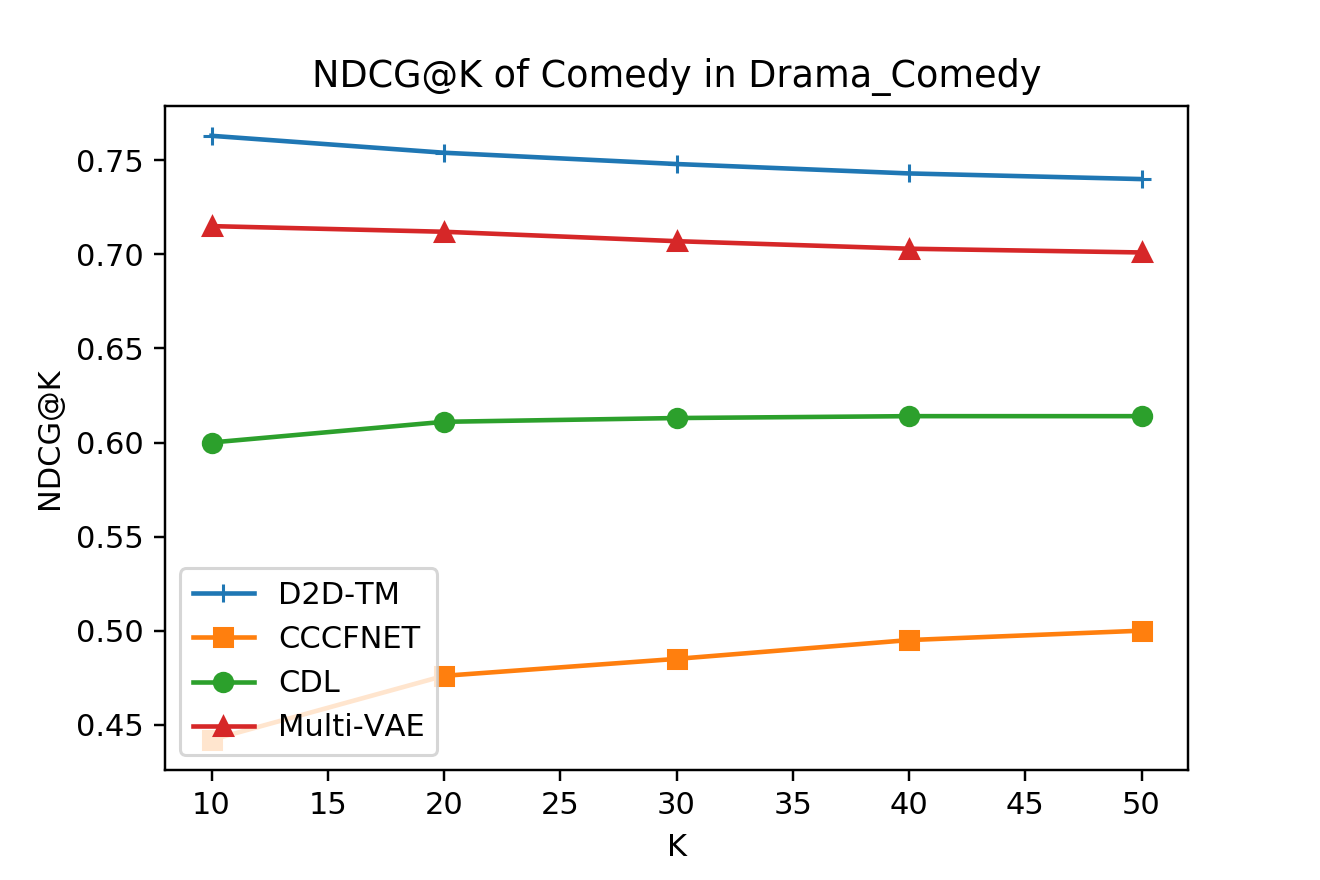}} \\
\subfloat[]{\includegraphics[width = 1.7in]{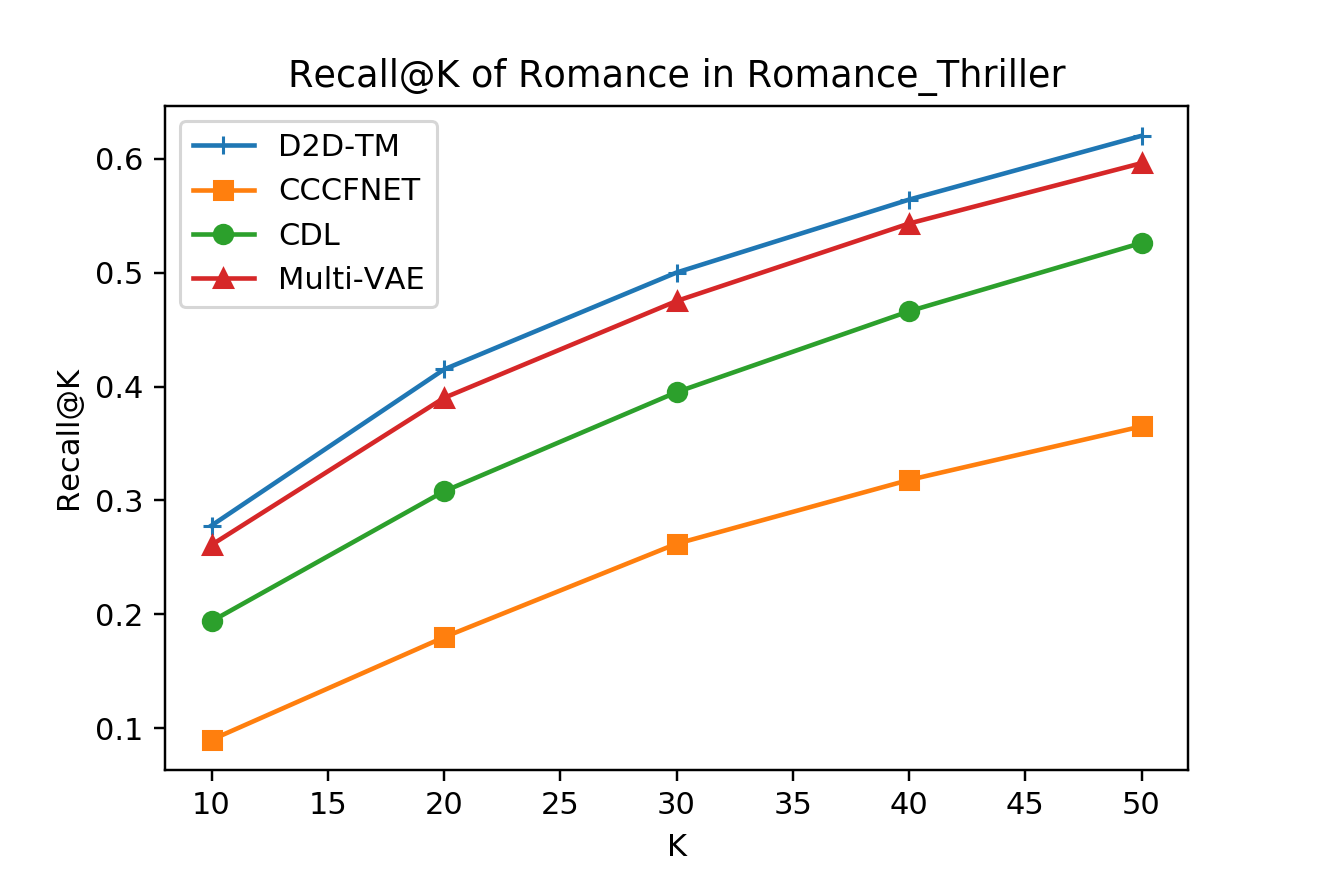}}
\subfloat[]{\includegraphics[width = 1.7in]{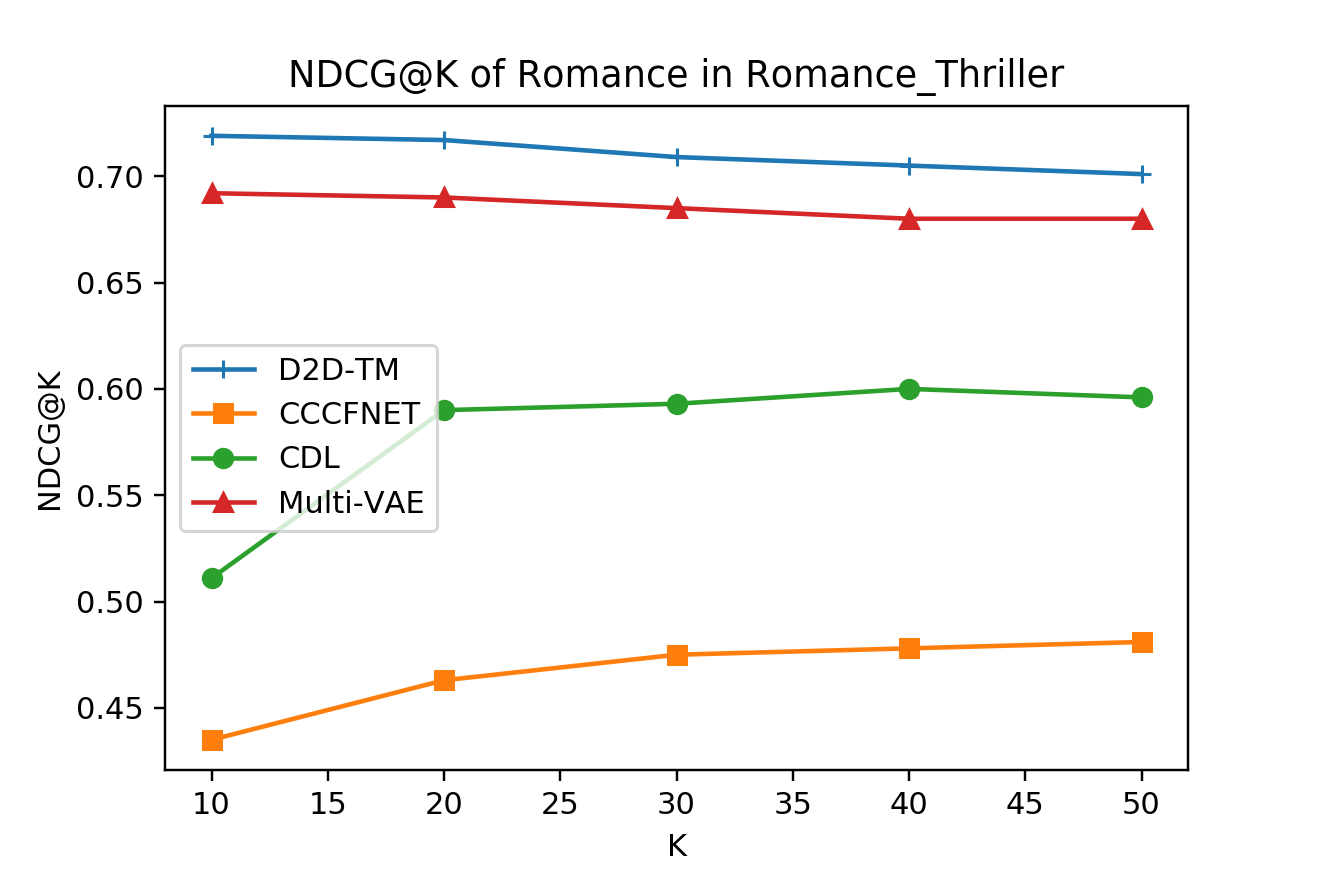}}
\subfloat[]{\includegraphics[width = 1.7in]{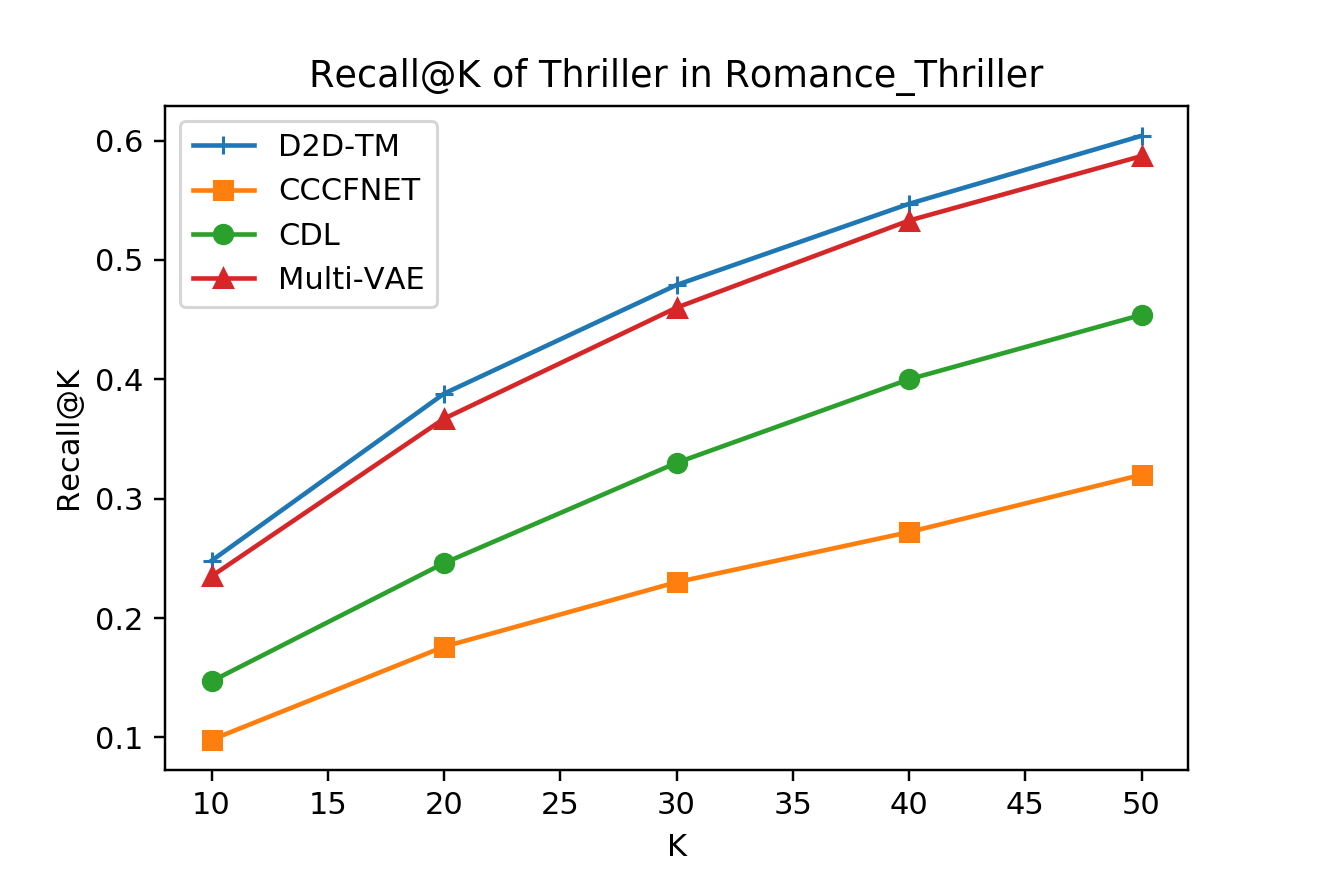}}
\subfloat[]{\includegraphics[width = 1.7in]{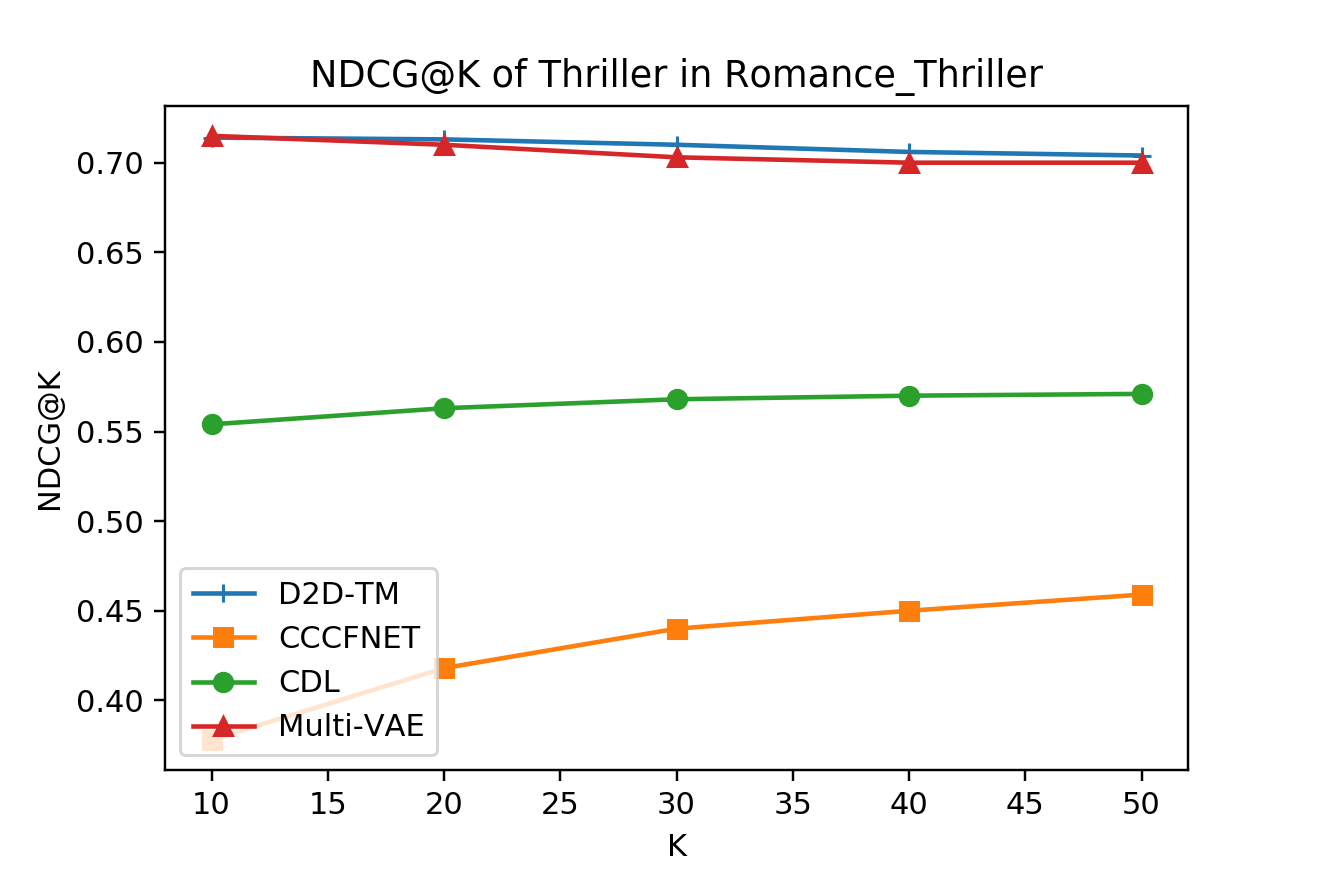}}
\caption{Recall and NDCG in four datasets with four methods.}
\label{result}
\end{figure*}

The models included in our comparison are listed as follows:
\begin{itemize}
\item \textbf{CDL}: Collaborative Deep Learning \cite{Wang2015} is a probabilistic feedforward model for the joint learning of stacked denoising autoencoder (SDAE) and collaborative filtering. For item contents, we combined the title and descriptions in Health\_Clothing and Video\_TV datasets as well as crawler movie descriptions from the IMBD website \footnote{https://www.imdb.com/} for Drama\_Comedy and Romance\_Thriller datasets. Then we merged products of the two domains into one set. Subsequently, we followed the same procedure as that explained in \cite{Wang2015} to preprocess the text information. After removing stop words, the top discriminate words according to the tf-idf values were chosen to form the vocabulary. We chose 8000 words for each dataset. Next, we use grid search and the validation set to ascertain the optimal hyperparamters. We search $\lambda_u$ in [0.1,1,10], $\lambda_v$ in [1, 10, 100] and $\lambda_r$ in [0.1, 1, 10]. Results demonstrated that the two-layer model which has detailed architecture as '8000-200-50-200-8000' yielded the best results in validation sets.
\item \textbf{Multi-VAE}: Multi-VAE \cite{Liang2018} is a collaborative filtering method that uses Variational Autoencoder (VAE) to reconstruct a user--item rating matrix. We concatenated two user--item matrixes from two domains so that the click vector of user $u$ is $[x_{1A}, x_{2A}, \cdots, x_{IA}, x_{1B}, \cdots, x_{IB}]$. Results demonstrated that structure '\#products-600-200-50-200-600-\#products' with a dimension of latent vector 50 yielded superior results in validation sets.
\item \textbf{CCCFNET}: Content-Boosted Collaborative Filtering Neural Network \cite{Lian2017} is a state-of-the-art hybrid method for cross-domain recommender systems. With a user, it uses a one-hot-encoding vector that extracts from a user--item rating matrix, but with the item, it combines both one-hot-encoding vector from user--item matrix and item attributes. Then, after learning, user hidden representation will include Collaborative Filtering (CF) factors and content preference, whereas item hidden representation includes CF factors and item content representation. We combine text attributes as in CDL with a user--item matrix, so that with each domain, the item input vector is $[x_{u1}, x_{u2}, \cdots, x_{uN}, x_{w1}, x_{w2}, \cdots, x_{wS}]$ for which N is the  number of users and S is 8000. The best neural network structure is '200-50'.
\item \textbf{APR}: Adversarial Personal Ranking \cite{He2018} enhances Bayesian Personal Ranking with an adversarial network. We use publicly available source code provided by authors, but it cannot obtain competitive performance on the datasets used for this study. Therefore, we do not plot the results of APR in Figure \ref{result}
\end{itemize}

\begin{figure}
\subfloat[]{\includegraphics[width = 1.6in]{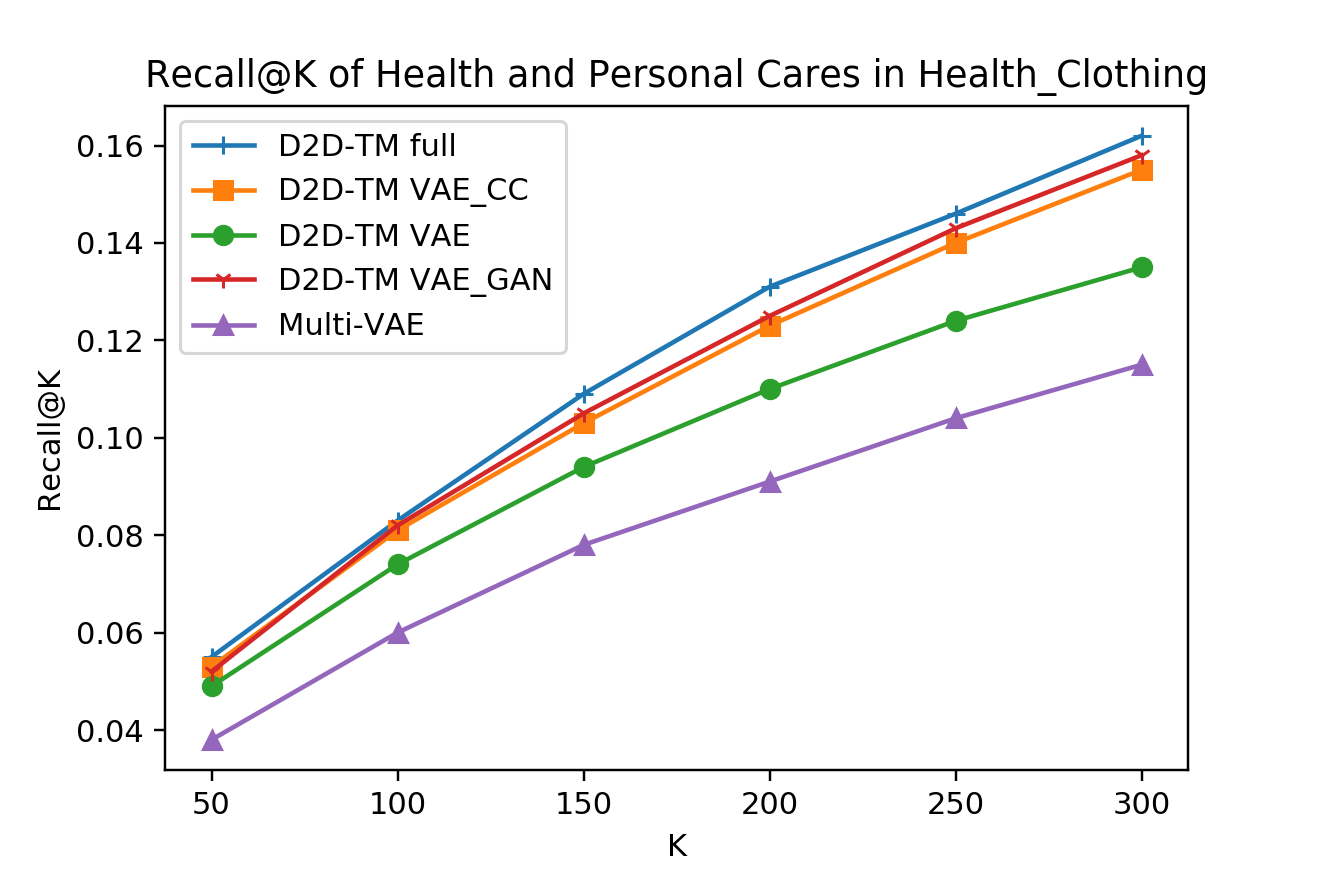}} 
\subfloat[]{\includegraphics[width = 1.6in]{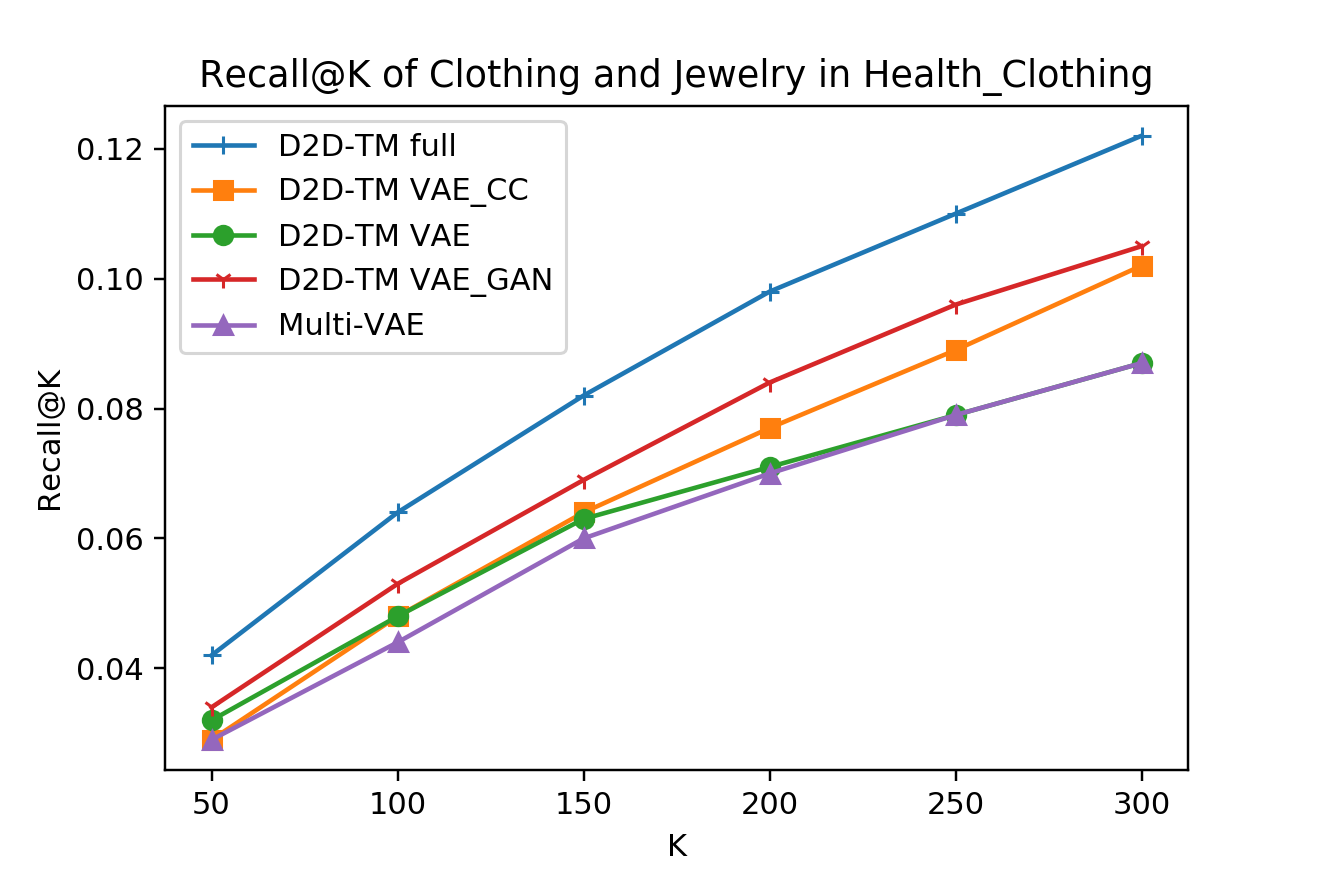}}
\caption{Comparing the recall of model components in the Health\_Clothing dataset.}
\label{component}
\end{figure}

\begin{figure}
\subfloat[]{\includegraphics[width = 1.6in]{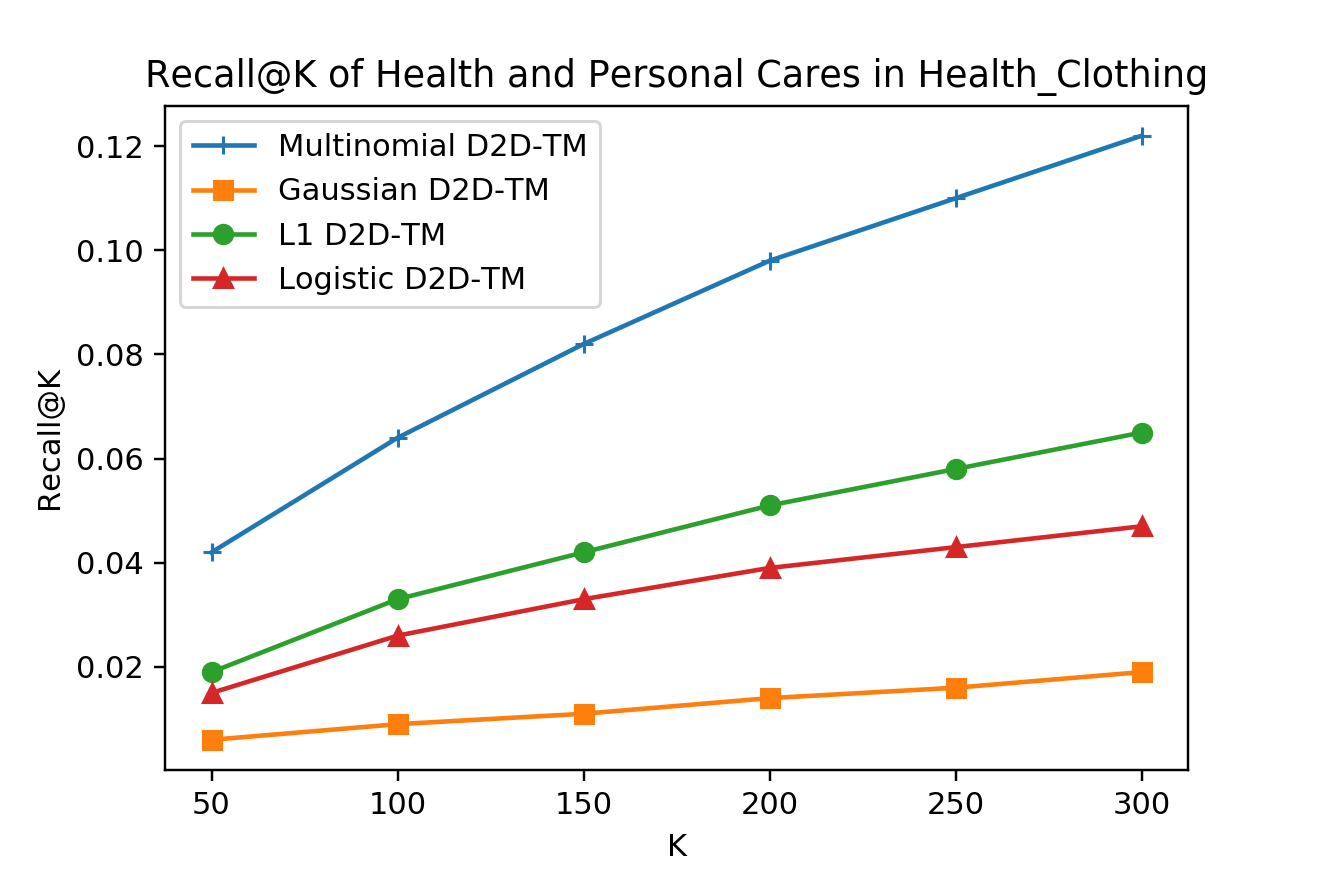}} 
\subfloat[]{\includegraphics[width = 1.6in]{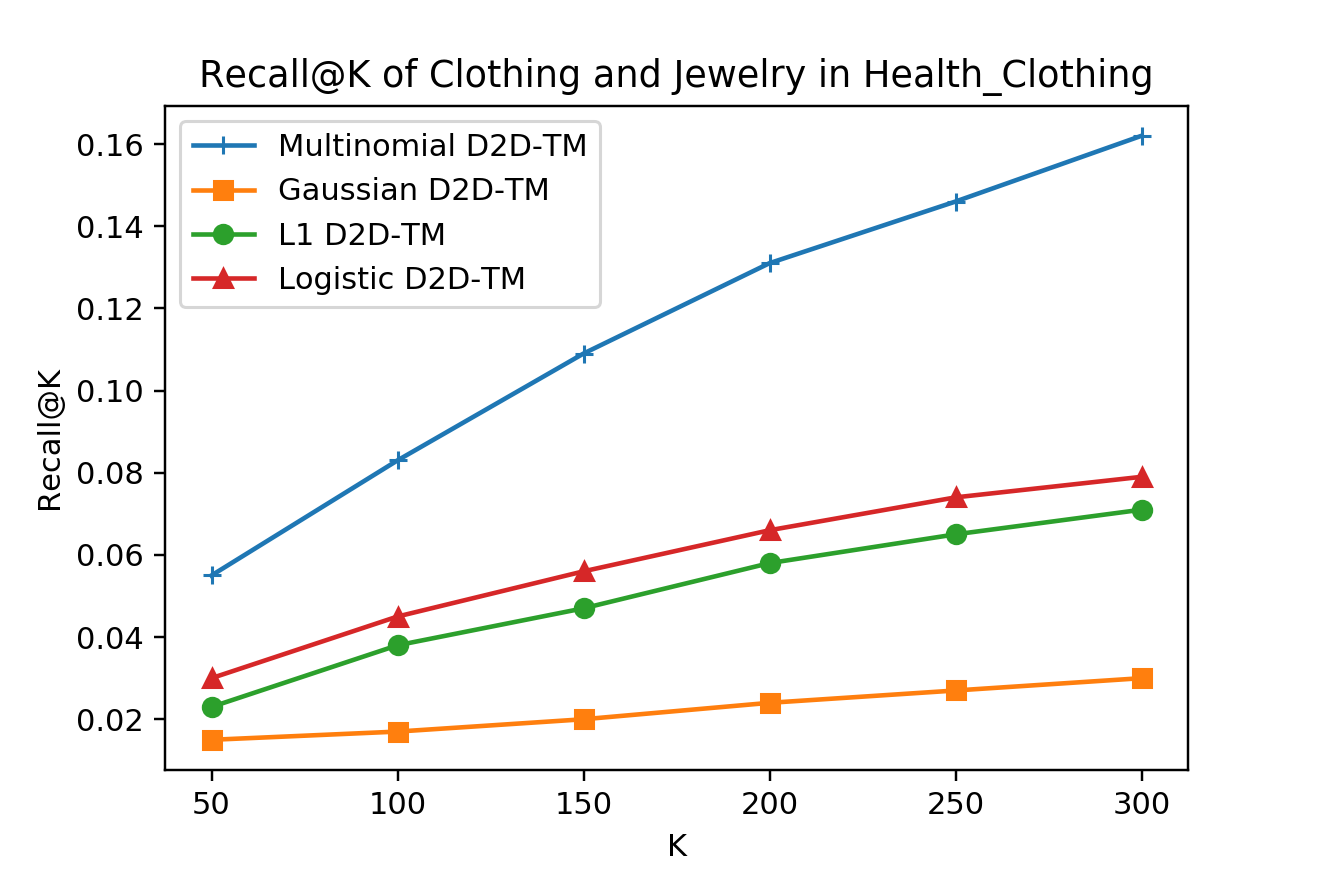}}
\caption{Compare recall of reconstruction loss functions for the Health\_Clothing dataset.}
\label{loss}
\end{figure}

\subsection{Performance Comparison}

%
%

Figure \ref{result} presents Recall and NDCG results of Multi-VAE, CDL, CCCFNET, and D2D-TM of each domain in four datasets. In light of these results, we have made the following observations.
\begin{itemize}
\item With \textbf{Multi-VAE}, it has some similar characteristics with our model such as using user interaction vectors as input and learning features through VAE. The main difference is that our model can learn differences between domains. It is apparent that if two domains differ in a certain attribute (Romance\_Thriller and Drama\_Comedy dataset), our model is higher than Multi-VAE obly 2.9\%--7.8\% in Recall@50. However, with two domains that differ in many attributes such as Health, Personal Care, and Clothing, Jewelry in the Health\_Clothing dataset, our model outperforms Multi-VAE with 44.8\% in Recall@50. Another reason is that only VAE might let the system overfit while extracting features by VAE. In such cases, discriminating by GAN helps the system avoid overfitting. Moreover, it can learn latent features better. The result demonstrates that learning specific features of each domain and integrating VAE-GAN can enhance performance.
\item With \textbf{CDL}, although it is a hybrid method, our model still can outperform by 17.9\% (Thriller) to 129\% (Health) in Recall@50. The first reason is similar to that for Multi-VAE: single-domain methods do not work well in multiple domains. The second reason is that different with CDL, our model only needs to train some users who have many interactions in both domains, but it can infer for all users. It not only reduces sparsity problems; it is also appropriate with real systems in cases where we need not do retraining when a new user comes.
\item With \textbf{APR}, we are unable to obtain competitive performance. In addition to the same reasons given for Multi-VAE and CDL, another possible reason is that GAN might work well in generating problems but not in extracting features as VAE. In our model, VAE is the main model to learn features, and the purpose of GAN is supporting for VAE in obtaining good features of two domains by trying to distinct generations between them. 
\item  With \textbf{CCCFNET}, a hybrid cross-domain method, our model can outperform by 52.7\% (Health) to 88.8\% (Thriller) in Recall@50. A possible reason is that the VAE-GAN model can learn latent features better than the simple Multilayer perceptron model can.
\end{itemize}

All four algorithms in baselines worked with the assumption that a user's behavior does not change. Even with CCCFNET, the user behavior is modeled as a sole network. However, based on special characteristics of each domain, user behavior presents some differences among domains. For example, a user is a saver who only bought inexpensive clothes, but with health care products, he must follow a doctor's advice and might purchase based on effectiveness, not on price. Our model has ability to capture both similar and different features of user behavior. Therefore, it is reasonable that our model can outperform the baselines.

Figure \ref{component} and Figure \ref{loss} respectively present the effectiveness of each component in our model as well as result of multinomial likelihood.

\subsubsection{Component}
Because VAE is key model to learn latent feature, we can keep VAE and try to ignore CC, GAN, or both. We designate D2D-TM full, D2D-TM VAE\_CC, D2D-TM VAE\_GAN, and D2D-TM VAE respectively as our original model, model ignoring CC, ignoring GAN and ignoring both CC and GAN. Experiments in Figure \ref{component} show that both CC and GAN are important to achieve high performance. However, results of D2D-TM VAE\_GAN are slightly better than those for D2D-TM VAE\_CC. A possible result is that GAN creates a strong constraint to avoid VAE overfitting so that VAE can extract latent features better. 

Weight-sharing and CC are important parts by which similarity can be learned between two domains, shown as D2D-TM VAE\_CC is higher than D2D-TM VAE 8.1\% in Health and Personal Care.  

The result that D2D-TM VAE is slightly better than Multi-VAE also demonstrates that learning different domains separately can improve performance.
\subsubsection{Reconstruction Loss Function}
In the UNIT framework, they use L1 loss for reconstruction. That is suitable with image data, but with click data, Multinomial log loss is more appropriate. Otherwise, many studies about RS used log likelihood (log loss) or Gaussian likelihood (square loss). Therefore, we experimented with loss of four types. With L1 loss, log loss and square loss, the activation function tanh can achieve superior results.

Figure \ref{loss} shows that the Multinomial log likelihood can outperform other types. A possible reason is that with click dataset, each element in the input vector is 0 or 1. Therefore, the square loss and L1 loss are unsuitable. Otherwise, the click input is assumed to be generated from a multinomial distribution. Demonstrably, it is better than log likelihood.

\section{Conclusion}
This paper presents a proposal of the D2D-TM network structure that can extract both homogeneous and divergent features among domains. This is first model reported to apply VAE-GAN to multi-domain RS. Moreover, our network can infer items in both domains simultaneously. Experiments have demonstrated that our proposed model can significantly outperform state-of-the-art methods for recommendation with more robust performance. Moreover, because our network uses only implicit feedback, it can be easily adopted for use by many companies.

%% file: finalmain.bbl

\begin{thebibliography}{21}


\ifx \showCODEN    \undefined \def \showCODEN     #1{\unskip}     \fi
\ifx \showDOI      \undefined \def \showDOI       #1{#1}\fi
\ifx \showISBNx    \undefined \def \showISBNx     #1{\unskip}     \fi
\ifx \showISBNxiii \undefined \def \showISBNxiii  #1{\unskip}     \fi
\ifx \showISSN     \undefined \def \showISSN      #1{\unskip}     \fi
\ifx \showLCCN     \undefined \def \showLCCN      #1{\unskip}     \fi
\ifx \shownote     \undefined \def \shownote      #1{#1}          \fi
\ifx \showarticletitle \undefined \def \showarticletitle #1{#1}   \fi
\ifx \showURL      \undefined \def \showURL       {\relax}        \fi
\providecommand\bibfield[2]{#2}
\providecommand\bibinfo[2]{#2}
\providecommand\natexlab[1]{#1}
\providecommand\showeprint[2][]{arXiv:#2}

\bibitem[\protect\citeauthoryear{Cai, Han, and Yang}{Cai et~al\mbox{.}}{2018}]%
        {Cai2018}
\bibfield{author}{\bibinfo{person}{Xiaoyan Cai}, \bibinfo{person}{Junwei Han},
  {and} \bibinfo{person}{Libin Yang}.} \bibinfo{year}{2018}\natexlab{}.
\newblock \showarticletitle{Generative Adversarial Network Based Heterogeneous
  Bibliographic Network Representation for Personalized Citation
  Recommendation}. In \bibinfo{booktitle}{\emph{AAAI}}.
\newblock


\bibitem[\protect\citeauthoryear{Cantador, Fern{\'a}ndez-Tob{\'\i}as,
  Berkovsky, and Cremonesi}{Cantador et~al\mbox{.}}{2015}]%
        {cantador2015cross}
\bibfield{author}{\bibinfo{person}{Iv{\'a}n Cantador}, \bibinfo{person}{Ignacio
  Fern{\'a}ndez-Tob{\'\i}as}, \bibinfo{person}{Shlomo Berkovsky}, {and}
  \bibinfo{person}{Paolo Cremonesi}.} \bibinfo{year}{2015}\natexlab{}.
\newblock \showarticletitle{Cross-domain recommender systems}.
\newblock In \bibinfo{booktitle}{\emph{Recommender Systems Handbook}}.
  \bibinfo{publisher}{Springer}, \bibinfo{pages}{919--959}.
\newblock


\bibitem[\protect\citeauthoryear{Cao and Yu}{Cao and Yu}{2016}]%
        {cao2016joint}
\bibfield{author}{\bibinfo{person}{Xuezhi Cao} {and} \bibinfo{person}{Yong
  Yu}.} \bibinfo{year}{2016}\natexlab{}.
\newblock \showarticletitle{Joint user modeling across aligned heterogeneous
  sites}. In \bibinfo{booktitle}{\emph{Proceedings of the 10th ACM Conference
  on Recommender Systems}}. ACM, \bibinfo{pages}{83--90}.
\newblock


\bibitem[\protect\citeauthoryear{Chen, Zhang, He, Nie, Liu, and Chua}{Chen
  et~al\mbox{.}}{2017}]%
        {chen2017attentive}
\bibfield{author}{\bibinfo{person}{Jingyuan Chen}, \bibinfo{person}{Hanwang
  Zhang}, \bibinfo{person}{Xiangnan He}, \bibinfo{person}{Liqiang Nie},
  \bibinfo{person}{Wei Liu}, {and} \bibinfo{person}{Tat-Seng Chua}.}
  \bibinfo{year}{2017}\natexlab{}.
\newblock \showarticletitle{Attentive collaborative filtering: Multimedia
  recommendation with item-and component-level attention}. In
  \bibinfo{booktitle}{\emph{Proceedings of the 40th International ACM SIGIR
  conference on Research and Development in Information Retrieval}}. ACM,
  \bibinfo{pages}{335--344}.
\newblock


\bibitem[\protect\citeauthoryear{Chu and Tsai}{Chu and Tsai}{2017}]%
        {Chu2017}
\bibfield{author}{\bibinfo{person}{Wei-Ta Chu} {and} \bibinfo{person}{Ya-Lun
  Tsai}.} \bibinfo{year}{2017}\natexlab{}.
\newblock \showarticletitle{A Hybrid Recommendation System Considering Visual
  Information for Predicting Favorite Restaurants}.
\newblock \bibinfo{journal}{\emph{World Wide Web}} \bibinfo{volume}{20},
  \bibinfo{number}{6} (\bibinfo{date}{Nov.} \bibinfo{year}{2017}),
  \bibinfo{pages}{1313--1331}.
\newblock
\showISSN{1386-145X}
\urldef\tempurl%
\url{https://doi.org/10.1007/s11280-017-0437-1}
\showDOI{\tempurl}


\bibitem[\protect\citeauthoryear{Elkahky, Song, and He}{Elkahky
  et~al\mbox{.}}{2015}]%
        {Elkahky2015}
\bibfield{author}{\bibinfo{person}{Ali~Mamdouh Elkahky}, \bibinfo{person}{Yang
  Song}, {and} \bibinfo{person}{Xiaodong He}.} \bibinfo{year}{2015}\natexlab{}.
\newblock \showarticletitle{A Multi-View Deep Learning Approach for Cross
  Domain User Modeling in Recommendation Systems}. In
  \bibinfo{booktitle}{\emph{Proceedings of the 24th International Conference on
  World Wide Web}} \emph{(\bibinfo{series}{WWW '15})}.
  \bibinfo{publisher}{International World Wide Web Conferences Steering
  Committee}, \bibinfo{address}{Republic and Canton of Geneva, Switzerland},
  \bibinfo{pages}{278--288}.
\newblock
\showISBNx{978-1-4503-3469-3}
\urldef\tempurl%
\url{https://doi.org/10.1145/2736277.2741667}
\showDOI{\tempurl}


\bibitem[\protect\citeauthoryear{Fern{\'a}ndez-Tob{\'\i}as, Cantador,
  Kaminskas, and Ricci}{Fern{\'a}ndez-Tob{\'\i}as et~al\mbox{.}}{2012}]%
        {fernandez2012}
\bibfield{author}{\bibinfo{person}{Ignacio Fern{\'a}ndez-Tob{\'\i}as},
  \bibinfo{person}{Iv{\'a}n Cantador}, \bibinfo{person}{Marius Kaminskas},
  {and} \bibinfo{person}{Francesco Ricci}.} \bibinfo{year}{2012}\natexlab{}.
\newblock \showarticletitle{Cross-domain recommender systems: A survey of the
  state of the art}. In \bibinfo{booktitle}{\emph{Spanish Conference on
  Information Retrieval}}. sn, \bibinfo{pages}{24}.
\newblock


\bibitem[\protect\citeauthoryear{Gong and Zhang}{Gong and Zhang}{2016}]%
        {Gong2016}
\bibfield{author}{\bibinfo{person}{Yuyun Gong} {and} \bibinfo{person}{Qi
  Zhang}.} \bibinfo{year}{2016}\natexlab{}.
\newblock \showarticletitle{Hashtag Recommendation Using Attention-based
  Convolutional Neural Network}. In \bibinfo{booktitle}{\emph{Proceedings of
  the Twenty-Fifth International Joint Conference on Artificial Intelligence}}
  \emph{(\bibinfo{series}{IJCAI'16})}. \bibinfo{publisher}{AAAI Press},
  \bibinfo{pages}{2782--2788}.
\newblock
\showISBNx{978-1-57735-770-4}
\urldef\tempurl%
\url{http://dl.acm.org/citation.cfm?id=3060832.3061010}
\showURL{%
\tempurl}


\bibitem[\protect\citeauthoryear{He, He, Du, and Chua}{He
  et~al\mbox{.}}{2018}]%
        {He2018}
\bibfield{author}{\bibinfo{person}{Xiangnan He}, \bibinfo{person}{Zhankui He},
  \bibinfo{person}{Xiaoyu Du}, {and} \bibinfo{person}{Tat-Seng Chua}.}
  \bibinfo{year}{2018}\natexlab{}.
\newblock \showarticletitle{Adversarial Personalized Ranking for
  Recommendation}. In \bibinfo{booktitle}{\emph{The 41st International ACM
  SIGIR Conference on Research \&\#38; Development in Information Retrieval}}
  \emph{(\bibinfo{series}{SIGIR '18})}. \bibinfo{publisher}{ACM},
  \bibinfo{address}{New York, NY, USA}, \bibinfo{pages}{355--364}.
\newblock
\showISBNx{978-1-4503-5657-2}
\urldef\tempurl%
\url{https://doi.org/10.1145/3209978.3209981}
\showDOI{\tempurl}


\bibitem[\protect\citeauthoryear{Hu, Cao, Xu, Cao, Gu, and Zhu}{Hu
  et~al\mbox{.}}{2013}]%
        {hu2013personalized}
\bibfield{author}{\bibinfo{person}{Liang Hu}, \bibinfo{person}{Jian Cao},
  \bibinfo{person}{Guandong Xu}, \bibinfo{person}{Longbing Cao},
  \bibinfo{person}{Zhiping Gu}, {and} \bibinfo{person}{Can Zhu}.}
  \bibinfo{year}{2013}\natexlab{}.
\newblock \showarticletitle{Personalized recommendation via cross-domain
  triadic factorization}. In \bibinfo{booktitle}{\emph{Proceedings of the 22nd
  international conference on World Wide Web}}. ACM, \bibinfo{pages}{595--606}.
\newblock


\bibitem[\protect\citeauthoryear{Li and She}{Li and She}{2017}]%
        {Li2017}
\bibfield{author}{\bibinfo{person}{Xiaopeng Li} {and} \bibinfo{person}{James
  She}.} \bibinfo{year}{2017}\natexlab{}.
\newblock \showarticletitle{Collaborative Variational Autoencoder for
  Recommender Systems}. In \bibinfo{booktitle}{\emph{Proceedings of the 23rd
  ACM SIGKDD International Conference on Knowledge Discovery and Data Mining}}
  \emph{(\bibinfo{series}{KDD '17})}. \bibinfo{publisher}{ACM},
  \bibinfo{address}{New York, NY, USA}, \bibinfo{pages}{305--314}.
\newblock
\showISBNx{978-1-4503-4887-4}
\urldef\tempurl%
\url{https://doi.org/10.1145/3097983.3098077}
\showDOI{\tempurl}


\bibitem[\protect\citeauthoryear{Lian, Zhang, Xie, and Sun}{Lian
  et~al\mbox{.}}{2017}]%
        {Lian2017}
\bibfield{author}{\bibinfo{person}{Jianxun Lian}, \bibinfo{person}{Fuzheng
  Zhang}, \bibinfo{person}{Xing Xie}, {and} \bibinfo{person}{Guangzhong Sun}.}
  \bibinfo{year}{2017}\natexlab{}.
\newblock \showarticletitle{CCCFNet: A Content-Boosted Collaborative Filtering
  Neural Network for Cross Domain Recommender Systems}. In
  \bibinfo{booktitle}{\emph{Proceedings of the 26th International Conference on
  World Wide Web Companion}} \emph{(\bibinfo{series}{WWW '17 Companion})}.
  \bibinfo{publisher}{International World Wide Web Conferences Steering
  Committee}, \bibinfo{address}{Republic and Canton of Geneva, Switzerland},
  \bibinfo{pages}{817--818}.
\newblock
\showISBNx{978-1-4503-4914-7}
\urldef\tempurl%
\url{https://doi.org/10.1145/3041021.3054207}
\showDOI{\tempurl}


\bibitem[\protect\citeauthoryear{Liang, Krishnan, Hoffman, and Jebara}{Liang
  et~al\mbox{.}}{2018}]%
        {Liang2018}
\bibfield{author}{\bibinfo{person}{Dawen Liang}, \bibinfo{person}{Rahul~G.
  Krishnan}, \bibinfo{person}{Matthew~D. Hoffman}, {and} \bibinfo{person}{Tony
  Jebara}.} \bibinfo{year}{2018}\natexlab{}.
\newblock \showarticletitle{Variational Autoencoders for Collaborative
  Filtering}. In \bibinfo{booktitle}{\emph{Proceedings of the 2018 World Wide
  Web Conference}} \emph{(\bibinfo{series}{WWW '18})}.
  \bibinfo{publisher}{International World Wide Web Conferences Steering
  Committee}, \bibinfo{address}{Republic and Canton of Geneva, Switzerland},
  \bibinfo{pages}{689--698}.
\newblock
\showISBNx{978-1-4503-5639-8}
\urldef\tempurl%
\url{https://doi.org/10.1145/3178876.3186150}
\showDOI{\tempurl}


\bibitem[\protect\citeauthoryear{Liu, Breuel, and Kautz}{Liu
  et~al\mbox{.}}{2017}]%
        {Ming-Yu2017}
\bibfield{author}{\bibinfo{person}{Ming-Yu Liu}, \bibinfo{person}{Thomas
  Breuel}, {and} \bibinfo{person}{Jan Kautz}.} \bibinfo{year}{2017}\natexlab{}.
\newblock \showarticletitle{Unsupervised Image-to-Image Translation Networks}.
\newblock In \bibinfo{booktitle}{\emph{Advances in Neural Information
  Processing Systems 30}}, \bibfield{editor}{\bibinfo{person}{I.~Guyon},
  \bibinfo{person}{U.~V. Luxburg}, \bibinfo{person}{S.~Bengio},
  \bibinfo{person}{H.~Wallach}, \bibinfo{person}{R.~Fergus},
  \bibinfo{person}{S.~Vishwanathan}, {and} \bibinfo{person}{R.~Garnett}}
  (Eds.). \bibinfo{publisher}{Curran Associates, Inc.},
  \bibinfo{pages}{700--708}.
\newblock
\urldef\tempurl%
\url{http://papers.nips.cc/paper/6672-unsupervised-image-to-image-translation-networks.pdf}
\showURL{%
\tempurl}


\bibitem[\protect\citeauthoryear{Min, Bao, Xu, Hossain, et~al\mbox{.}}{Min
  et~al\mbox{.}}{2015}]%
        {min2015cross}
\bibfield{author}{\bibinfo{person}{Weiqing Min}, \bibinfo{person}{Bing-Kun
  Bao}, \bibinfo{person}{Changsheng Xu}, \bibinfo{person}{M~Shamim Hossain},
  {et~al\mbox{.}}} \bibinfo{year}{2015}\natexlab{}.
\newblock \showarticletitle{Cross-Platform Multi-Modal Topic Modeling for
  Personalized Inter-Platform Recommendation.}
\newblock \bibinfo{journal}{\emph{IEEE Trans. Multimedia}}
  \bibinfo{volume}{17}, \bibinfo{number}{10} (\bibinfo{year}{2015}),
  \bibinfo{pages}{1787--1801}.
\newblock


\bibitem[\protect\citeauthoryear{Pan, Xia, Liu, Peng, and Ming}{Pan
  et~al\mbox{.}}{2016}]%
        {pan2016mixed}
\bibfield{author}{\bibinfo{person}{Weike Pan}, \bibinfo{person}{Shanchuan Xia},
  \bibinfo{person}{Zhuode Liu}, \bibinfo{person}{Xiaogang Peng}, {and}
  \bibinfo{person}{Zhong Ming}.} \bibinfo{year}{2016}\natexlab{}.
\newblock \showarticletitle{Mixed factorization for collaborative
  recommendation with heterogeneous explicit feedbacks}.
\newblock \bibinfo{journal}{\emph{Information Sciences}}  \bibinfo{volume}{332}
  (\bibinfo{year}{2016}), \bibinfo{pages}{84--93}.
\newblock


\bibitem[\protect\citeauthoryear{Shapira, Rokach, and Freilikhman}{Shapira
  et~al\mbox{.}}{2013}]%
        {shapira2013facebook}
\bibfield{author}{\bibinfo{person}{Bracha Shapira}, \bibinfo{person}{Lior
  Rokach}, {and} \bibinfo{person}{Shirley Freilikhman}.}
  \bibinfo{year}{2013}\natexlab{}.
\newblock \showarticletitle{Facebook single and cross domain data for
  recommendation systems}.
\newblock \bibinfo{journal}{\emph{User Modeling and User-Adapted Interaction}}
  \bibinfo{volume}{23}, \bibinfo{number}{2-3} (\bibinfo{year}{2013}),
  \bibinfo{pages}{211--247}.
\newblock


\bibitem[\protect\citeauthoryear{Wang, Wang, and Yeung}{Wang
  et~al\mbox{.}}{2015}]%
        {Wang2015}
\bibfield{author}{\bibinfo{person}{Hao Wang}, \bibinfo{person}{Naiyan Wang},
  {and} \bibinfo{person}{Dit-Yan Yeung}.} \bibinfo{year}{2015}\natexlab{}.
\newblock \showarticletitle{Collaborative Deep Learning for Recommender
  Systems}. In \bibinfo{booktitle}{\emph{Proceedings of the 21th ACM SIGKDD
  International Conference on Knowledge Discovery and Data Mining}}
  \emph{(\bibinfo{series}{KDD '15})}. \bibinfo{publisher}{ACM},
  \bibinfo{address}{New York, NY, USA}, \bibinfo{pages}{1235--1244}.
\newblock
\showISBNx{978-1-4503-3664-2}
\urldef\tempurl%
\url{https://doi.org/10.1145/2783258.2783273}
\showDOI{\tempurl}


\bibitem[\protect\citeauthoryear{Wang, Yu, Zhang, Gong, Xu, Wang, Zhang, and
  Zhang}{Wang et~al\mbox{.}}{2017b}]%
        {Wang2017}
\bibfield{author}{\bibinfo{person}{Jun Wang}, \bibinfo{person}{Lantao Yu},
  \bibinfo{person}{Weinan Zhang}, \bibinfo{person}{Yu Gong},
  \bibinfo{person}{Yinghui Xu}, \bibinfo{person}{Benyou Wang},
  \bibinfo{person}{Peng Zhang}, {and} \bibinfo{person}{Dell Zhang}.}
  \bibinfo{year}{2017}\natexlab{b}.
\newblock \showarticletitle{IRGAN: A Minimax Game for Unifying Generative and
  Discriminative Information Retrieval Models}. In
  \bibinfo{booktitle}{\emph{Proceedings of the 40th International ACM SIGIR
  Conference on Research and Development in Information Retrieval}}
  \emph{(\bibinfo{series}{SIGIR '17})}. \bibinfo{publisher}{ACM},
  \bibinfo{address}{New York, NY, USA}, \bibinfo{pages}{515--524}.
\newblock
\showISBNx{978-1-4503-5022-8}
\urldef\tempurl%
\url{https://doi.org/10.1145/3077136.3080786}
\showDOI{\tempurl}


\bibitem[\protect\citeauthoryear{Wang, He, Nie, and Chua}{Wang
  et~al\mbox{.}}{2017a}]%
        {Wang2017CD}
\bibfield{author}{\bibinfo{person}{Xiang Wang}, \bibinfo{person}{Xiangnan He},
  \bibinfo{person}{Liqiang Nie}, {and} \bibinfo{person}{Tat-Seng Chua}.}
  \bibinfo{year}{2017}\natexlab{a}.
\newblock \showarticletitle{Item Silk Road: Recommending Items from Information
  Domains to Social Users}. In \bibinfo{booktitle}{\emph{Proceedings of the
  40th International ACM SIGIR Conference on Research and Development in
  Information Retrieval}} \emph{(\bibinfo{series}{SIGIR '17})}.
  \bibinfo{publisher}{ACM}, \bibinfo{address}{New York, NY, USA},
  \bibinfo{pages}{185--194}.
\newblock
\showISBNx{978-1-4503-5022-8}
\urldef\tempurl%
\url{https://doi.org/10.1145/3077136.3080771}
\showDOI{\tempurl}


\bibitem[\protect\citeauthoryear{Wang, Wang, Li, He, and Liu}{Wang
  et~al\mbox{.}}{2013}]%
        {wang2013theoretical}
\bibfield{author}{\bibinfo{person}{Yining Wang}, \bibinfo{person}{Liwei Wang},
  \bibinfo{person}{Yuanzhi Li}, \bibinfo{person}{Di He}, {and}
  \bibinfo{person}{Tie-Yan Liu}.} \bibinfo{year}{2013}\natexlab{}.
\newblock \showarticletitle{A theoretical analysis of NDCG type ranking
  measures}. In \bibinfo{booktitle}{\emph{Conference on Learning Theory}}.
  \bibinfo{pages}{25--54}.
\newblock


\end{thebibliography}
